\documentclass[letterpaper,10pt,journal,twoside]{IEEEtran}

\usepackage{amsmath,mathtools,amssymb}
\allowdisplaybreaks
\usepackage{stmaryrd}
\usepackage{paralist}
\usepackage{bbold}
\usepackage{dsfont}
\usepackage{tikz,pgfplots,subfigure}
\usepackage[american]{circuitikz}
\usepackage{anyfontsize}
\pgfplotsset{compat=1.18} 
\usetikzlibrary{calc,arrows,shapes,backgrounds,calc,positioning,patterns,decorations.pathmorphing,decorations.markings,mindmap,trees}
\usepackage{blox}
\usepackage{color}
\usepackage{siunitx}	
\usepackage{booktabs}	
\usepackage{balance}
\usepackage{xcolor}
\usepackage{cite}
\graphicspath{{../figs/}}

\newtheorem{theorem}{Theorem}

\newtheorem{corollary}[theorem]{Corollary}
\newtheorem{proposition}[theorem]{Proposition}
\newtheorem{property}{Property}
\newtheorem{remark}{Remark}[section]

\newcounter{cAss}
\newcounter{cAssSaved}
\newcommand\Ass[1]{\ensuremath{\boldsymbol{\mathcal A}_{\text{\hspace{0.75pt}\bf#1}}}}
\newlength\asswidth
 {\begin{list}{\Ass{\arabic{cAss}}:}{%
   \setcounter{cAssSaved}{\thecAss}\usecounter{cAss}\setcounter{cAss}{\thecAssSaved}
   \setlength\topsep{.667ex plus 2pt minus 2pt}
   \setlength\itemsep\topsep
   \settowidth\asswidth{\Ass{\arabic{cAss}}:}
   \addtolength\asswidth\labelsep
   \setlength\leftmargin\asswidth
   \setlength\labelwidth\asswidth}}%
 {\end{list}}

\makeatletter
\xdef\@endgadget#1{{\unskip\nobreak\hfil\penalty50\hskip1em\hbox{}\nobreak\hfil#1\parfillskip=0pt\finalhyphendemerits=0\par}}
\newcommand\@Endofsymbol{$\triangledown$}
\newcommand\Endofremark{\@endgadget{\@Endofsymbol}}
\makeatother

\newcommand{\R}{\mathbb{R}}
\newcommand{\N}{\mathbb{N}}
\newcommand{\ee}{\mathrm{e}}
\newcommand{\dd}{\mathrm{d}}
\newcommand{\xe}{x_{\text{e}}}

\DeclareMathOperator*{\argmin}{argmin}

\begin{document}
	\title{Stabilization of Switched Affine Systems\\ With Dwell-Time Constraint}
	\author{Antonio Russo, \IEEEmembership{Member, IEEE}, Gian Paolo Incremona, \IEEEmembership{Senior Member, IEEE}\\ and Patrizio Colaneri, \IEEEmembership{Fellow, IEEE}
	\thanks{This work was supported by the Italian Ministry for Research (PRIN 2022, PRIDE, grant no. 2022LP77J4 and contract FSE-REACT-EU, PON Ricerca e Innovazione 2014-2020 DM1062/2021), and by the selection notice issued by the Rector's decree n. 509 of 13/06/2022 for the funding of fundamental and applied research projects dedicated to young researchers - ACTNOW.}
	\thanks{Antonio Russo is with Dipartimento di Ingegneria, Universit\`a degli Studi della Campania ``L. Vanvitelli'', 81031 Aversa, Italy (e-mail: \textsl{antonio.russo1@unicampania.it}).}
     \thanks{Gian Paolo Incremona and Patrizio Colaneri are with Dipartimento di Elettronica, Informazione e Bioingegneria, Politecnico di Milano, 20133 Milan, Italy (e-mails: 	\textsl{gianpaolo.incremona@polimi.it}, \textsl{patrizio.colaneri@polimi.it}).}}
	
	\maketitle
	
	\begin{abstract}
This paper addresses the problem of stabilization of switched affine systems under dwell-time constraint, giving guarantees on the bound of the quadratic cost associated with the proposed state switching control law. Specifically, two switching rules are presented relying on the solution of  differential Lyapunov inequalities and  Lyapunov-Metzler inequalities, from which the stability conditions are expressed.
The first one allows to regulate the state of  linear switched systems to zero, whereas the second one is designed for switched affine systems proving practical stability of the origin. In both cases, the determination of a guaranteed cost associated with each control strategy is shown. In the cases of linear and affine systems, the existence of the solution for the Lyapunov-Metzler condition is discussed and guidelines for the selection of a solution ensuring suitable performance of the system evolution are provided. The theoretical results are finally assessed by means of three examples.
	\end{abstract}
	

	
\section{Introduction}
\IEEEPARstart{S}{witched} systems are a pivotal class characterized by the interplay of continuous and discontinuous dynamics. These systems consist of multiple subsystems, each governed by distinct dynamical equations, and a switching signal that manages the transitions between them \cite{switching:Liberzon:2003,switching:Sun:2005}. In the literature, it is well established that switched systems do not inherently possess the stability properties of their subsystems. Therefore, the primary challenges in studying switched systems are investigating their stability under various classes of switching signals \cite{switched:LiberzonMorse:1999} and defining an effective switching strategy to achieve stabilization \cite{LM:GeromelColaneri:2006}.

In the context of system stabilization, it is important to distinguish between stabilization by time-driven switching signal and by state-feedback switching signal. It is not surprising to understand that there are systems that cannot be stabilized using time-driven switching signal but  are stabilizable through a well-designed state-feedback strategy \cite{LM:GeromelColaneri:2006}. A notable exception is given by the class of positive, more in general monotone, switched systems \cite{positive:Blanchini:2015}. A further characterization of the set of possible stabilizing switching signals is given by the concept of (hard or average) dwell-time switching \cite{stability:Zhai:2000,stability:DellaRossa:2023}, that limits the frequency of the commutation signal among the system's modes. Without this limit, the commutation can be arbitrarily fast, and the associated state-dependent feedback law can be designed through the so-called Lyapunov-Metzler inequalities, as in \cite{LM:GeromelColaneri:2006}.
A review of results on stability and stabilization of switched linear systems can be found in \cite{survey:Lin:2009,detstoc:Colaneri:2009}.

\subsection{State of the art for switched affine systems}
As an extension of switched linear systems, switched affine ones are used to model many processes characterized by variable structures, such as power converters \cite{powerconverter:Albea:2021}, traffic congestion models \cite{traffic:Blanchini:2012}, biochemical networks \cite{bio:Parise:2018} and robotic locomotion \cite{multiple:Alpcan:2010}. Unlike switched linear systems, switched affine systems can have different equilibrium points for each subsystem, and the control objective is to design a switching law that regulates the state to a point outside the set of subsystems' equilibria.

Designing stabilizing strategies for switched affine systems remains an open and challenging problem. For example, the pioneering work \cite{arbitrarySwitching:Bolzern:2004} demonstrates that the existence of a convex combination of subsystems forming a linear Hurwitz average system is sufficient to establish a switching strategy for regulating the state to the desired equilibrium. An optimal control strategy for switched piecewise affine autonomous systems, aimed at minimizing a performance index over an infinite time horizon, is proposed in \cite{optimalaffine:Seatzu:2006}. A local stability result is proposed in \cite{local:Hetel:2015}, while a robust sampled-data approach was pursued in \cite{sampled:Hetel:2013}. Furthermore, in \cite{disturbed:Kader:2018}, a state-dependent switching law is developed to stabilize switched affine systems, even when state measurements are affected by perturbations and noise. When the subsystems' equilibria are unknown, \cite{unknownequilibrium:Deaecto:2022} provides an approach based on a two-steps procedure expressed in terms of linear matrix inequalities (LMIs). Additionally, \cite{rankdeficient:Egidio:2022} addresses the global stabilization of continuous-time switched affine systems with rank-deficient convex combinations of their dynamic matrices. Finally, the recent paper \cite{datadriven:Seuret:2023} proposes a systematic method to translate a model-based condition, expressed using the framework of LMIs, into a data-driven condition. However, these control strategies permit potentially arbitrarily fast switching.

While arbitrarily fast switching can generate average solutions in the Filippov sense \cite{differential:Filippov:1988}, dwell-time switching techniques are necessary for systems that require a finite commutation frequency, due to physical limitations, or to prevent actuator wear and tear. Conversely, while arbitrarily fast switching allows global asymptotic stabilization of the desired equilibrium point, the dwell-time constraint prevents the asymptotic stabilization to this desired equilibrium (unless it coincides with the equilibrium point of one of the subsystems). Various strategies have been proposed to practically stabilize switched affine systems under a dwell-time constraint. For instance, \cite{affine:AlbeaZaccarian:2019} proposes a state-dependent switching control strategy that ensures a dwell-time constraint by using time-regularization within a hybrid system framework. In \cite{switched:Egidio:2020}, a dwell-time constraint is inherently enforced by guaranteeing global asymptotic stability of a limit cycle for affine switched systems. Similarly, \cite{practical:Xu:2021} analyzes the stability of both continuous-time and discrete-time switched affine systems via dwell-time switching. This approach constructs a discretized Lyapunov function to provide sufficient conditions to ensure practical stability of the state trajectories. Finally, \cite{timeevent:Albea:2021} designs periodic time- and event-triggered control laws for switched affine systems using a hybrid dynamical system approach. This work, along with an appropriate optimization problem, formulates a stabilization result to ensure uniform global asymptotic stability of an attractor near the desired operating point for both types of controllers. 

\subsection{Contributions with respect to the state of the art}
The present work, first, inspired by \cite{dwellTime:Allerhand:2013},  introduces a novel dwell-time switching strategy aimed at regulating the state of a linear switched system to the origin. This proposed switching law utilizes a time-varying Lyapunov function derived from the solution of a differential Lyapunov inequality and Lyapunov-Metzler inequalities. A key innovation of our approach compared to existing literature is its ability to provide a guaranteed bound on the $\mathcal{H}_2$ cost function. This bound aligns with those found in the literature regarding the control of switched linear systems with arbitrarily fast switching \cite{LM:GeromelColaneri:2006}.

Then, we generalize the proposed switching strategy to switched affine systems, ensuring practical stability of the origin and, differently from our previous findings \cite{affine:Russo:2022}, a guaranteed bound on the average $\mathcal{H}_2$ cost function, contingent on the dwell time. It is demonstrated that this bound on the average $\mathcal{H}_2$ cost function can be determined, in a simplified setting, as the solution to an optimization problem based on LMI and a line search procedure. Unlike other works in the literature, such as \cite{arbitrarySwitching:Bolzern:2004,affine:AlbeaZaccarian:2019,timeevent:Albea:2021}, our proposed switching strategies do not depend on the existence of a convex combination of subsystems that generates a Hurwitz average linear system. Therefore, the proposed switching law can be successfully applied to switched affine systems that do not admit such linear Hurwitz convex combination, e.g., the traffic congestion control problem presented in \cite{traffic:Blanchini:2012}.

\subsection{Structure of the paper}
The paper is organized as follows. In Section \ref{sec:setup}, the problem of stabilization of switched affine systems is formalized. First, the design of a dwell-time switching law for switched linear systems is discussed in depth in Section \ref{sec:switching_linear}. Then, the extension to the case of switched affine systems is introduced and rigorously analyzed in Section \ref{sec:switching_affine}. Practices on the design of the switching law tuning parameters are provided in Section \ref{sec:tuning}. Finally, the effectiveness of the proposed algorithms is demonstrated through three examples in Section \ref{sec:examples}, while some conclusions are gathered in Section \ref{sec:conclusions}.

\medskip%
\subsubsection*{Notation}
The transpose of a matrix $A$ is denoted by $A'$. The set of reals is notated as $\R$, while the sets of non-negative real and natural numbers are $\R_{\geq 0}$ and $\N_{\geq 0}$, respectively. Signals in the time domains are denoted by lower-case letters, like $x(t)$, or just $x$.  We indicated with $\mathcal{M}$ the class of irreducible Metzler matrices, which consist of all matrices $\Pi \in \R^{M \times M}$ with elements $\pi_{i,j}$ such that $\pi_{i,j}\geq 0$ for all $i \neq j$, and 
$\sum_{j=1}^M \pi_{i,j} = 0$, for all $i\in \{1,\dots, M\}$.
For Hermitian matrices, $X>0$ (resp. $X\geq 0$) indicates that $X$ is positive (resp. semi-positive) definite. For given matrices $A$ and $B$ of compatible dimensions, $A\otimes B$ indicates the Kronecker product of the two matrices, whereas the Kronecker sum is defined as $A\oplus B = A\otimes I + I \otimes B$, with $I$ being the identity matrix. Finally, the symbols $0_n$ and $1_n$ indicate the $n$-dimensional column vectors with all zero entries and all one entries, respectively. Similarly, $0_{n\times n}$ indicates the $n$ by $n$ null matrix. 

\section{Problem Setup}\label{sec:setup}
Consider the switched affine system 
 \begin{subequations}\label{eq:affine_sys}
     \begin{align} 
		\dot{x}(t) &= A_{\sigma(t)}x(t)+b_{\sigma(t)},\quad x(t_0)=x_0 \label{eq:sys}\\
		z(t)  &= Cx(t),\label{eq:z}
	\end{align}
 \end{subequations}
where the state $x\in\R^n$ is available for feedback for all $t\geq 0$, $z\in\R^p$ is a performance output and $x_0$ is the initial condition. Let $\Omega\coloneqq \{1,\dots, M\}$, and denote with $\Omega_i$ the set of integers $\{1,\dots,i-1,i+1,\dots,M\}$, with $M\in\N_{>0}$. Then, considering a set of matrices $A_i \in \R^{n\times n}$ and vectors $b_i \in \R^{n}$, $i \in \Omega$, be given, the switching law $\sigma(t)$, for each $t \geq 0$ is such that
\begin{equation*}
    A_{\sigma(t)}\in\{A_1,\dots,A_M\},\ \quad b_{\sigma(t)}\in \{b_1, \dots, b_M\},
\end{equation*}
with $\{t_k\},\,k=0,1,\dots,\infty$ being the monotonically increasing sequence of time instants such that $A_{\sigma(t_k^-)} \neq A_{\sigma(t_k)}$ or $b_{\sigma(t_k^-)} \neq b_{\sigma(t_k)}$. In what follows, we will say that the switching signal $\sigma(t)$ has dwell time $T$ if $t_{k+1}-t_k\geq T>0$ for all $k=0,1,\dots,\infty$, with $t_0$ being the initial time instant. The control objective is that of designing a switching law capable of stabilizing an operating point $\xe\in\R^n$, and without loss of generality, in the following we consider the case where the switched equilibrium coincides with the origin, that is $\xe=0$, since any prescribed equilibrium point can be shifted to the origin via a change of variable $\tilde{x}=x-\xe$.

The design of the switching law presented in this work takes inspiration from the stripped-down stabilization problem with dwell time of the linear switched system 
\begin{subequations}\label{eq:linear_sys}
    \begin{align}
        \dot{x}(t) &= A_{\sigma(t)}x(t),\quad x(t_0)=x_0 \\
		z(t)  &= Cx(t),
    \end{align}
\end{subequations}
presented in \cite{dwellTime:Allerhand:2013}. The idea of \cite{dwellTime:Allerhand:2013} is to design a switching dependent quadratic Lyapunov function taking into account the solution of the system dynamics within a time horizon equal to the dwell time $T$. 
Making reference to system \eqref{eq:linear_sys}, the approach in \cite[\S III, Thm. 1]{dwellTime:Allerhand:2013} may be presented as follows. Assuming that there exists a collection $\{P_1, \dots, P_M\}$ of positive definite matrices, and $\Pi \in \mathcal{M}$, such that the following Lyapunov-Metlzer inequalities hold,
\begin{equation}
    P_iA_i+A_i'P_i+\sum_{j\in\Omega_i} \pi_{i,j}\left(\!\ee^{A_j'T}P_j\ee^{A_j T}-P_i\!\right)\!<\!0, \enskip\forall\,i\!\in\!\Omega,
\end{equation}
assuming $\sigma(t_k)=i$, then the switching law is defined as
 \begin{align}
    \sigma(t) & = i, \enskip  \forall t \in [t_k, t_k+T] \nonumber  \\
	\sigma(t) & = i,  \enskip  \forall t >t_k+T, \nonumber \\
	       &\textrm{if}\enskip x(t)' P_i x(t) \geq  x(t)'\ee^{A_j'T}P_j\ee^{A_jT} x(t), \quad  \forall j \in \Omega_i \nonumber  \\
	\sigma(t_{k+1}) &= \argmin_{j\in\Omega_i}x(t_{k+1})'\ee^{A_j'T}P_j\ee^{A_jT} x(t_{k+1}), \nonumber \\
                    &\text{otherwise.} \label{eq:Shaked_Allerhand}
 \end{align}
 Then, it is proved in \cite[\S III, Thm. 1]{dwellTime:Allerhand:2013} that the switching law \eqref{eq:Shaked_Allerhand} globally asymptotically stabilizes the system $\dot{x}(t) = A_{\sigma(t)}x(t)$ with dwell-time constraint $T$.
 Conceptually, the law in \eqref{eq:Shaked_Allerhand} compares the current value of the Lyapunov function $V(x,t)=x'(t)P_{\sigma(t)}x(t)$ of the active subsystem with the Lyapunov function corresponding to the other subsystems evaluated at $T$ time units ahead of the current time instant. It is worth to notice that the law \eqref{eq:Shaked_Allerhand}, although it does not limit the application to affine linear switched systems, is designed without taking into account the affine term $b_{\sigma(t)}$ and its effect on the performance index, commonly defined as
\begin{equation} \label{eq:cost}
         J(x_0,t) = \int_{t_0}^t z'(\tau)z(\tau) \dd\tau.
\end{equation}
The latter may indeed be considered as a measure of the efficiency of the algorithm and is of great importance when selecting the best approach in applications, providing a robustness indicator of the process behaviour.

The cost \eqref{eq:cost}  is not  easily manageable in the case of \cite{dwellTime:Allerhand:2013}, and the switching law \eqref{eq:Shaked_Allerhand} is not readily extendable to include the affine term.
Motivated by the above two challenges, we now present an alternative approach to provide some result on the cost, also including in the switching law the term $b_{\sigma}$. 

More specifically, we address the following problems:
\begin{enumerate}
    \item[\emph{i)}] \emph{design a state-feedback switching law $\sigma(t)= u(x(t),t)$ with dwell time $T$ that globally asymptotically stabilizes the origin of the switched linear system \eqref{eq:linear_sys}}, and provide a compatible bound to $J(x_0,t)$ for each $t\geq t_0$; 
    \item[\emph{ii)}] \emph{design a state-feedback switching law $\sigma(t)= u(x(t),t)$ with dwell time $T$ that globally practically stabilizes the origin of the switched affine system \eqref{eq:affine_sys}}, and provide a compatible bound to $J(x_0,t)$ for each $t\geq t_0$.
\end{enumerate}
     
\section{dwell-time State-Feedback Switching Law for\\
 Linear Switched Systems} \label{sec:switching_linear}
In this section, to solve the first problem previously formulated, we propose an enhanced state-feedback switching control strategy, providing, differently from \cite{dwellTime:Allerhand:2013}, an estimation on the performance cost \eqref{eq:cost}. 

Consider system \eqref{eq:linear_sys} and let us define, for each $j\in \Omega$, the following quantities,
\begin{subequations}
		\begin{align*}
			Y_{1,j} & = e^{A_j' T}X_je^{A_j T}, \enskip &Y_{2,j} & =\!\! \int_0^T e^{A_j'\tau}C'Ce^{A_j \tau}\mathrm{d}\tau,
		\end{align*}
\end{subequations}
where $X_j$ is the solution to a Lyapunov-Metzler inequality, introduced in the next theorem.
\begin{theorem}\label{thm:differential_P}
    Consider the switched linear system \eqref{eq:linear_sys} and assume that there exist positive definite symmetric time-varying matrices ${P_i(t) \colon \R_{\geq 0} \to \R^{n\times n}}$, constant symmetric positive definite matrices $X_i \in \R^{n\times n}$ and $\Pi \in \mathcal{M}$ so that
\begin{subequations}\label{eq:strat1_Pdot}
\begin{align}
        -\dot{P}_i \!&=\! A_i' P_i \!+\! P_i A_i \!+\! C' C,  &&\forall t \!\in\! [t_k,t_k\!+\!T), \forall i \in \Omega \label{eq:strat1_Pdot_1} \\
        P_i(t) \!& =\! X_i,  && \forall t \geq t_k+T, \forall i \in \Omega \label{eq:strat1_Pdot_2}
\end{align}
\end{subequations}
with matrices $X_i$ being the solution of the Lyapunov-Metzler inequalities
\begin{equation}\label{eq:strat1_LyapMetz}
        A_i' X_i + X_i A_i + \sum\limits_{j \in \Omega_i} \pi_{i,j}(Y_{1,j}+Y_{2,j} - X_i) + C' C < 0, 
\end{equation}
for any $i \in \Omega$. Then, the dwell-time switching strategy
\begin{subequations}\label{eq:strat1}
\begin{align}
        \sigma(t) & = i, \enskip  \forall t \in [t_k, t_k+T]  \label{eq:strat1_cond0} \\
        \sigma(t) & = i,  \enskip  \forall t >t_k+T, \label{eq:strat1_cond2} \\
        &\textrm{if}\enskip x(t)' (Y_{1,j}+Y_{2,j}) x(t) \geq  x(t)'X_i x(t), \quad  \forall j \in \Omega_i  \nonumber \\
        \sigma(t_{k+1}) &= \argmin_{j\in\Omega_i}x(t_{k+1})'(Y_{1,j}+Y_{2,j}) x(t_{k+1}) \label{eq:strat1_cond3}
\end{align}
\end{subequations}
where 
\begin{equation}\label{eq:tkp1}
        t_{k+1}\coloneqq\!\! \inf_{t > t_k+T}\! \left\lbrace t | \exists j \colon x(t)'\left[ Y_{1,j}\!+\!Y_{2,j}-X_i\right]x(t)\!<\!0 \right\rbrace,
\end{equation}
 guarantees that the origin of system \eqref{eq:linear_sys} is globally exponentially stable. Furthermore, the performance index \eqref{eq:cost} is bounded as 
\begin{equation}\label{eq:strat1_cost}
     J(x_0,t) \leq x(t_0)'P_{\sigma(t_0)}(t_0)x(t_0).
\end{equation}
\end{theorem}

\begin{IEEEproof}
Let us consider the following Lyapunov function,
\begin{equation}
    V(x,t) = \begin{cases}
        x(t)'P_{\sigma(t)}(t)x(t) \quad &t \in [t_k, t_k+T) \\
        x(t)'X_{\sigma(t)}x(t) \quad &t \in [t_k+T, t_{k+1}), \label{eq:strat1_V}
    \end{cases}
\end{equation}
where $t_{k+1}$ is defined as in \eqref{eq:tkp1}. Note that this function is continuous, by construction, at time instant $t_k+T$. Initially, consider the interval $[t_k, t_k+T)$, with the subsystem $i$ being active, i.e., $\sigma(t)=i$, for all $t\in [t_k, t_k+T)$. Then, from \eqref{eq:strat1_V}, $V(x,t) = x(t)'P_i(t)x(t)$ and
\begin{align}
    \dot{V}(x,t) & = \dot{x}(t)'P_i(t)x(t) + x(t)'\dot{P}_i(t)x(t)+ x(t)'P_i(t)\dot{x}(t) \nonumber\\
    & = x(t)'(A_i'P_i(t)+P_i(t)A_i+\dot{P}_i)x(t) \nonumber \\
    & = -x(t)'C'Cx(t), \label{eq:strat1_step1}
\end{align}
where the last equality follows from \eqref{eq:strat1_Pdot_1}.

Consider now the interval $[t_k+T, t_{k+1})$. Since switching has not been triggered yet, then $\sigma(t)=i$ for all $t\in [t_k+T, t_{k+1})$. Hence, from \eqref{eq:strat1_V}, $V(x,t) = x(t)'X_ix(t)$ and
\begin{align}
    \dot{V}(x,t)  =& \dot{x}(t)'X_ix(t) + x(t)'X_i\dot{x}(t) \nonumber\\
     =& x(t)'(A_i'X_i+X_iA_i)x(t) \nonumber \\
     \leq& - x(t)'\left[\sum\limits_{j \in \Omega_i} \pi_{i,j}(Y_{1,j}+Y_{2,j} - X_i) + C' C\right]x(t) \nonumber \\
     \leq& -x(t)'C'Cx(t), \label{eq:strat1_step2}
\end{align}
where the first inequality comes from \eqref{eq:strat1_LyapMetz}, while the last inequality comes from \eqref{eq:strat1_cond2} in the switching rule. 

Let us now consider the jumps of the Lyapunov function at the switching instants, i.e.,
\begin{align*}
    \Delta &V(x(t_{k+1}),t_{k+1}) = V(x(t_{k+1}),t_{k+1})-V(x(t_{k+1}^-),t_{k+1}^-) \nonumber \\
    & = x(t_{k+1})'P_j(t_{k+1})x(t_{k+1})-x(t_{k+1})'P_i(t_{k+1}^-)x(t_{k+1}).
\end{align*}
Note that from \eqref{eq:strat1_Pdot_1} 
\begin{align*}
    P_j(t_{k+1}) & = e^{A_j' T}X_je^{A_j T}+ \int_{0}^{T} e^{A_j' \tau}C' C e^{A_j \tau} \mathrm{d}\tau\nonumber \\
    & = Y_{1,j}+Y_{2,j}\,, \\
    P_i(t_{k+1}^-) & = X_i\,,
\end{align*}   
which implies
\begin{multline}\label{eq:deltaV}
\Delta V(x(t_{k+1}),t_{k+1}) =  \\
  x(t_{k+1})'(Y_{1,j}+Y_{2,j}-X_i)x(t_{k+1})<0\,, 
\end{multline}
where the inequality comes from considering the switching law condition \eqref{eq:strat1_cond2}.

Thus, from \eqref{eq:strat1_step1}, \eqref{eq:strat1_step2} and \eqref{eq:deltaV} one has that $\dot{V}(x,t) \leq -x(t)'C'Cx(t)$ for almost all $t \in [t_k, t_{k+1}]$ and $\Delta V(x(t_{k+1}),t_{k+1})<0$ at the switching instant. Hence, iterating the above steps results in $\dot{V}(x,t) \leq -x(t)'C'Cx(t)$ for almost all $t\geq t_0$, thus implying global exponential stability of the origin. Finally, integrating both sides it holds that
\begin{align*}
    J(x_0,t) = \int_{t_0}^{t} z(\tau)'z(\tau) \mathrm{d}\tau &\leq V(x(t_0),t_0)\\
    &=x(t_0)'P_{\sigma(t_0)}(t_0)x(t_0),
\end{align*}
thus proving the theorem statement.
\end{IEEEproof}

The previous theorem shows that the switching law \eqref{eq:strat1_LyapMetz} globally asymptotically stabilizes the origin of the switched linear system \eqref{eq:linear_sys} while providing a guaranteed upper-bound for the $\mathcal{H}_2$ cost. 
\begin{remark}[Applicability of the switching control law]\label{rem:applicability}
    It is worth emphasizing that requiring the existence of matrices $P_i(t)$ such that conditions \eqref{eq:strat1_Pdot} are satisfied does not require matrices $A_i$ to be Hurwitz, since \eqref{eq:strat1_Pdot} is assumed to hold only within a finite time interval. \hfill$\triangledown$
\end{remark}
\begin{remark}[Initialization of the switching signal]\label{rem:initialization_linear}
    Different approaches are possible to initialize the switching signal $\sigma$. For instance, a convenient approach is to initialize the switching signal as $\sigma(t_0) = \argmin_{\sigma} V(x(t_0),t_0)$ to minimize the cost \eqref{eq:strat1_cost}. \hfill$\triangledown$
\end{remark}
An interesting result can be obtained if the switching strategy \eqref{eq:strat1} is straightforwardly applied to the switched affine system \eqref{eq:affine_sys}, as it is shown in the next corollary.
\begin{corollary}
    \label{corol:Pdot_extension}
    Consider the switched affine system \eqref{eq:affine_sys} and assume that there exist positive definite symmetric time-varying matrices ${P_i(t) \colon \R_{\geq 0} \to \R^{n\times n}}$, constant symmetric positive definite matrices $X_i \in \R^{n\times n}$, $\Pi \in \mathcal{M}$ and scalar $\epsilon>0$ so that
    \begin{subequations}\label{eq:corol_strat1_Pdot}
    \begin{align}
            -\dot{P}_i \!&=\! \hat{A}_i' P_i \!+\! P_i \hat{A}_i \!+\! C' C,  &&\forall t \!\in\! [t_k,t_k\!+\!T), \forall i \in \Omega \label{eq:corol_strat1_Pdot_1} \\
            P_i(t) \!& =\! X_i,  && \forall t \geq t_k+T, \forall i \in \Omega \label{eq:corol_strat1_Pdot_2}
    \end{align}
    \end{subequations}
    with $\hat{A}_i=A_i+\frac{\epsilon}{2}I$ and matrices $X_i$ being the solution of the Lyapunov-Metzler inequalities
    \begin{equation}\label{eq:corol_strat1_LyapMetz}
            \hat{A}_i' X_i + X_i \hat{A}_i + \sum\limits_{j \in \Omega_i} \pi_{i,j}(Y_{1,j}+Y_{2,j} - X_i) + C' C < 0, 
    \end{equation}
    for any $i \in \Omega$. Then, the dwell-time switching strategy \eqref{eq:strat1} guarantees that the origin of system \eqref{eq:affine_sys} is practically stable.    
    Furthermore, the cost is bounded as 
    \begin{align}\label{eq:corol_cost}
        J(x_0,t)=\int_{t_0}^{t} &z(\tau)'z(\tau) \mathrm{d}\tau \nonumber \\
        &\leq x(t_0)'P_{\sigma(t_0)}(t_0)x(t_0) + \delta(t-t_0),
    \end{align}
    where 
    \begin{equation}\label{eq:corol_delta}
        \delta = \frac{1}{\epsilon}\max_{i \in \Omega}\bigg\{\max_{t\in[t_k,t_k+T)} b_i'P_i(t)b_i\bigg\},
    \end{equation}
\end{corollary}
\begin{IEEEproof}
The proof follows similar reasoning as in Theorem \ref{thm:differential_P}, by selecting the Lyapunov function as in \eqref{eq:strat1_V}. Initially, consider the interval $[t_k, t_k+T)$, with the subsystem $i$ being active, i.e., $\sigma(t)=i$, for all $t\in [t_k, t_k+T)$. Then, computing the Lyapunov function time derivative within the interval $[t_k,t_k+T)$, the equality \eqref{eq:strat1_step1} becomes
    \begin{equation}
        \label{eq:corol_Vdot1}
        \dot{V}(x,t) = -x(t)'C'Cx(t) -\epsilon x(t)'P_i(t)x(t)+ 2b_i'P_i(t)x(t).
    \end{equation} 
    Considering the time interval $[t_k+T,t_{k+1})$, instead, the derivative of the Lyapunov function is upper-bounded as
    \begin{equation}\label{eq:corol_Vdot2}
        \dot{V}(x,t) \leq -x(t)'C'Cx(t)-\epsilon x(t)'X_ix(t)+ 2b_i'X_ix(t).
    \end{equation}
    Finally, the Lyapunov function difference at the switching instants still satisfies condition \eqref{eq:deltaV} due to the switching law definition.

   Hence, from \eqref{eq:corol_Vdot1} one obtains
	\begin{align}
		\dot{V}(x,t) & \leq -x(t)'C'C x(t) + \delta_i, \quad \forall t \in [t_k,t_{k}+T),
	\end{align}
	where $\delta_{i} \coloneqq \tfrac{1}{\epsilon}\max_{t\in[t_k,t_k+T)} b_i'P_i(t)b_i$. Similarly, from \eqref{eq:corol_Vdot2} one obtains
	\begin{align}
		\dot{V}(x,t) & \leq -x(t)'C'C x(t) + \frac{1}{\epsilon}b_i'X_ib_i \nonumber \\
		& \leq -x(t)'C'C x(t) + \delta_{i}, \quad \forall t \in [t_k+T,t_{k+1}),
	\end{align}
	where the last inequality derives from $\delta_i \geq \tfrac{1}{\epsilon}b_i'X_ib_i$ due to continuity of $P_i(t)$ at $t=t_k+T$. At the switching instants condition, \eqref{eq:deltaV} holds. Therefore, for almost all $t\in \R_{\geq 0}$, it holds that
	\begin{equation}
		\dot{V}(x,t) \leq - x(t)'C'C x(t) + \delta,
	\end{equation}
	with $\delta \coloneqq \max_{i\in \Omega}\{\delta_{i}\}$ and $\Delta V(x(t_k),t_k)<0$ for all $k$. Thus, practical stability is implied. Integrating both sides it holds that
	\begin{equation}
		\int_{t_0}^{t} z(\tau)'z(\tau) \mathrm{d}\tau \leq V(x(t_0),t_0) + \delta(t-t_0),
	\end{equation}
	thus obtaining \eqref{eq:corol_cost}.
\end{IEEEproof}
\begin{remark}[Relation with Theorem \ref{thm:differential_P}]
    The conditions specified in Theorem \ref{thm:differential_P} are precisely the ones required for Corollary \ref{corol:Pdot_extension} achieved by setting $\epsilon=0$. In other words, the prerequisites for both results are identical, except for the specific value of $\epsilon$. In Corollary \ref{corol:Pdot_extension}, it would be desirable to find the largest positive $\epsilon$ that satisfies conditions \eqref{eq:corol_strat1_Pdot} and \eqref{eq:corol_strat1_LyapMetz}, in order to minimize the cost upper-bound $\delta$. \hfill$\triangledown$
\end{remark}

In light of the proof of Theorem \ref{thm:differential_P}, we may solve only the Lyapunov-Metzeler inequality \eqref{eq:strat1_LyapMetz}, without computing the solution to the differential Lyapunov equation \eqref{eq:strat1_Pdot}, but just exploiting the definitions of $Y_{1,j}$ and $Y_{2,j}$. Moreover, it is worth noticing that the proposed solution provides a bound on the performance index depending on the system initial condition, in analogy with existing literature results on arbitrarily fast switching \cite{LM:GeromelColaneri:2006}. 
\begin{remark}[Average cost ultimate bound]
    From \eqref{eq:corol_cost} it is clear that, in the long run
    \begin{equation*}
        \lim_{t\to\infty} \frac{1}{t-t_0}\int_{t_0}^t z(\tau)'z(\tau) \mathrm{d}\tau \leq \delta\,.
    \end{equation*}\hfill$\triangledown$
\end{remark}
Finally, it must be noted that the switching law \eqref{eq:strat1} does not explicitly take into account the affine terms $b_i$. In the following, an alternative switching strategy accounting for the affine terms is presented.

\section{Practical Stabilization\\ for Switched Affine Systems} \label{sec:switching_affine}
Differently from the control law proposed in the previous section, it is now desirable to include an affine term to the evaluation of the switching condition. 
Therefore, an easy-to-implement switching strategy is hereafter presented, as a natural extension of the solution reported in Theorem \ref{thm:differential_P}. It is worth highlighting that, differently from Corollary \ref{corol:Pdot_extension}, the proposed switching strategy explicitly takes into account the affine term in the definition of the switching law.

Let us introduce the following extended system
 \begin{subequations}\label{eq:extended_affine_sys}
     \begin{align} 
		\dot{\tilde{x}}(t) &= \tilde{A}_{\sigma(t)}\tilde{x}(t),\quad \tilde{x}(t_0)=\tilde{x}_0 \label{eq:extended_sys}\\
		z(t)  &= \tilde{C}\tilde{x}(t),\label{eq:extended_z}
	\end{align}
 \end{subequations}
 with $\tilde{x} = [x' \enskip \bar{x}]'$, $\tilde{x}_0 = [x_0' \enskip 1]'$ and   
 \begin{equation}
    \tilde{A}_{\sigma(t)} = \begin{bmatrix}
        A_{\sigma(t)} & b_{\sigma(t)} \\
        0_n' & 0
    \end{bmatrix}, \quad \tilde{C} = [C \enskip 0_p].
 \end{equation}
 It is easy to see that system \eqref{eq:extended_affine_sys} is equivalent to system \eqref{eq:affine_sys}. Moreover, hereafter the following auxiliary variables are used
 \begin{subequations}
    \begin{align*}
        \tilde{Y}_{1,j} & = e^{\tilde{A}_j' T}\tilde{X}_je^{\tilde{A}_j T}, \enskip &\tilde{Y}_{2,j} & =\!\! \int_0^T e^{\tilde{A}_j'\tau}\tilde{C}'\tilde{C}e^{\tilde{A}_j \tau}\mathrm{d}\tau,
    \end{align*}
\end{subequations}
with
$\tilde{X}_j$ being the solution to a Lyapunov-Metzler inequality, introduced  in Theorem \ref{thm:main}, structured as
\begin{equation}\label{eq:tildeX_j}
    \tilde{X}_j = \begin{bmatrix}
        X_{j} & 0_n \\
        0_n' & 1
    \end{bmatrix},
\end{equation}
$X_j$ being symmetric and positive definite. Finally, let us define
\begin{equation*}
    \tilde{I} = 
        \begin{bmatrix}
            0_{n\times n} & 0_n \\
            0_n' & 1
        \end{bmatrix}.
\end{equation*}
In the following, the proposed dwell-time switching law for practical stabilization of the switched affine system \eqref{eq:affine_sys} is formulated.
\begin{theorem} \label{thm:main}
Consider the switched system \eqref{eq:extended_affine_sys} and assume that there exist positive definite symmetric time-varying matrices ${\tilde{P}_i(t) \colon \R_{\geq 0} \to \R^{(n+1)\times (n+1)}}$, constant symmetric positive definite matrices $\tilde{X}_i \in \R^{(n+1)\times (n+1)}$, $i \in \Omega$, structured as in \eqref{eq:tildeX_j}, $\Pi \in \mathcal{M}$ and scalar $\varepsilon>0$ so that
\begin{subequations}\label{eq:strat2_Pdot}
\begin{align}
        -\dot{\tilde{P}}_i \!&=\! \tilde{A}_i' \tilde{P}_i \!+\! \tilde{P}_i \tilde{A}_i \!+\! \tilde{C}' \tilde{C},  &&\forall t \!\in\! [t_k,t_k\!+\!T), \forall i \in \Omega \label{eq:strat2_Pdot_1} \\
        \tilde{P}_i(t) \!& =\! \tilde{X}_i,  && \forall t \geq t_k+T, \forall i \in \Omega \label{eq:strat2_Pdot_2}
\end{align}
\end{subequations}
with matrices $\tilde{X}_i$ being the solution of the Lyapunov-Metzler inequalities
\begin{equation}\label{eq:strat2_LyapMetz}
        \tilde{A}_i' \tilde{X}_i + \tilde{X}_i \tilde{A}_i + \sum\limits_{j \in \Omega_i} \pi_{i,j}(\tilde{Y}_{1,j}+\tilde{Y}_{2,j} - \tilde{X}_i) + \tilde{C}' \tilde{C} < \varepsilon \tilde{I},  
\end{equation}
Then, the dwell-time switching strategy
\begin{subequations}\label{eq:strat2}
\begin{align}
        \sigma(t) & = i, \enskip  \forall t \in [t_k, t_k+T]  \label{eq:strat2_cond0} \\
        \sigma(t) & = i,  \enskip  \forall t >t_k+T, \label{eq:strat2_cond2} \\
        &\textrm{if}\enskip \tilde{x}(t)' (\tilde{Y}_{1,j}+\tilde{Y}_{2,j}) \tilde{x}(t) \geq  \tilde{x}(t)'\tilde{X}_i \tilde{x}(t), \quad  \forall j \in \Omega_i  \nonumber \\
        \sigma(t_{k+1}) &= \argmin_{j\in\Omega_i}\tilde{x}(t_{k+1})'(\tilde{Y}_{1,j}+\tilde{Y}_{2,j}) \tilde{x}(t_{k+1}) \label{eq:strat2_cond3}
\end{align}
\end{subequations}
where 
\begin{equation}\label{eq:tkp1_ext}
        \!\!\!\!t_{k+1}\coloneqq\!\! \inf_{t > t_k+T}\! \left\lbrace t | \exists j \colon \tilde{x}(t)'\left[ \tilde{Y}_{1,j}\!+\!\tilde{Y}_{2,j}-\tilde{X}_i\right]\tilde{x}(t)\!<\!0 \right\rbrace,
\end{equation}
 guarantees that the origin of system \eqref{eq:affine_sys} is practically stable. Furthermore, the performance index \eqref{eq:cost} is bounded as 
\begin{equation}\label{eq:strat2_cost}
     J(x_0,t)=\int_{t_0}^{t} z(\tau)'z(\tau) \mathrm{d}\tau \leq  x(t_0)'P_{\sigma(t_0)}(t_0)x(t_0) + \varepsilon(t-t_0).
\end{equation}
\end{theorem}
 
\begin{IEEEproof}
    Let us consider the following Lyapunov function,
\begin{equation}
    V(\tilde{x},t) = \begin{cases}
        \tilde{x}(t)'\tilde{P}_{\sigma(t)}(t)\tilde{x}(t) \quad &t \in [t_k, t_k+T) \\
        \tilde{x}(t)'\tilde{X}_{\sigma(t)}\tilde{x}(t) \quad &t \in [t_k+T, t_{k+1}), \label{eq:strat2_V}
    \end{cases}
\end{equation}
where $t_{k+1}$ is defined as in \eqref{eq:tkp1_ext}. Note that this function is continuous, by construction, at time instant $t_k+T$. Initially, consider the interval $[t_k, t_k+T)$, with the subsystem $i$ being active, i.e., $\sigma(t)=i$, for all $t\in [t_k, t_k+T)$. Then, from \eqref{eq:strat2_V}, $V(\tilde{x},t) = \tilde{x}(t)'\tilde{P}_{i}(t)\tilde{x}(t)$ and
\begin{align}
    \dot{V}(\tilde{x},t) & = \dot{\tilde{x}}(t)'\tilde{P}_i(t)\tilde{x}(t) + \tilde{x}(t)'\dot{\tilde{P}}_i(t)\tilde{x}(t)+ \tilde{x}(t)'\tilde{P}_i(t)\dot{\tilde{x}}(t) \nonumber\\
    & = \tilde{x}(t)'(\tilde{A}_i'\tilde{P}_i(t)+\tilde{P}_i(t)\tilde{A}_i+\dot{\tilde{P}}_i)\tilde{x}(t) \nonumber \\
    & = -\tilde{x}(t)'\tilde{C}'\tilde{C}\tilde{x}(t) = -x(t)'C'Cx(t), \label{eq:strat2_step1}
\end{align}
where the third equality follows from \eqref{eq:strat2_Pdot_1}. 

Consider now the interval $[t_k+T, t_{k+1})$. Since switching has not been triggered yet, then $\sigma(t)=i$ for all $t\in [t_k+T, t_{k+1})$. Hence, from \eqref{eq:strat2_V}, $V(\tilde{x},t) = \tilde{x}(t)'\tilde{X}_i\tilde{x}(t)$ and
\begin{align}
    \dot{V}(&\tilde{x},t)  = \dot{\tilde{x}}(t)'\tilde{X}_i\tilde{x}(t) + \tilde{x}(t)'\tilde{X}_i\dot{\tilde{x}}(t) \nonumber\\
     =& \tilde{x}(t)'(\tilde{A}_i'\tilde{X}_i+\tilde{X}_i\tilde{A}_i)\tilde{x}(t) \nonumber \\
     \leq& - \tilde{x}(t)'\left[\sum\limits_{j \in \Omega_i} \pi_{i,j}(\tilde{Y}_{1,j}+\tilde{Y}_{2,j} - \tilde{X}_i) + \tilde{C}' \tilde{C} +\varepsilon \tilde{I}\right]\tilde{x}(t) \nonumber \\
     \leq& -\tilde{x}(t)'\tilde{C}' \tilde{C}\tilde{x}(t) +\varepsilon \tilde{x}(t)'\tilde{I}\tilde{x}(t) \nonumber \\
     = & -x(t)'C'Cx(t) + \varepsilon, \label{eq:strat2_step2}
\end{align}
where the first inequality comes from \eqref{eq:strat2_LyapMetz}, while the second inequality comes from \eqref{eq:strat2_cond2} in the switching rule. 

Let us now consider the jumps of the Lyapunov function at the switching instants, i.e.,
\begin{align*}
    \Delta &V(\tilde{x}(t_{k+1}),t_{k+1}) = V(\tilde{x}(t_{k+1}),t_{k+1})-V(\tilde{x}(t_{k+1}^-),t_{k+1}^-) \nonumber \\
    & = \tilde{x}(t_{k+1})'\tilde{P}_j(t_{k+1})\tilde{x}(t_{k+1})-\tilde{x}(t_{k+1})'\tilde{P}_i(t_{k+1}^-)\tilde{x}(t_{k+1}).
\end{align*}
Note that from \eqref{eq:strat2_Pdot_1} 
\begin{align*}
    \tilde{P}_j(t_{k+1}) & = e^{\tilde{A}_j' T}\tilde{X}_je^{\tilde{A}_j T}+ \int_{0}^{T} e^{\tilde{A}_j' \tau}\tilde{C}' \tilde{C} e^{\tilde{A}_j \tau} \mathrm{d}\tau\nonumber \\
    & = \tilde{Y}_{1,j}+\tilde{Y}_{2,j}\,, \\
    \tilde{P}_i(t_{k+1}^-) & = \tilde{X}_i\,,
\end{align*}   
which implies
\begin{align}\label{eq:strat2_deltaV}
\Delta V(\tilde{x}(t_{k+1}),t_{k+1}) &=  \tilde{x}(t_{k+1})'(\tilde{Y}_{1,j}+\tilde{Y}_{2,j}-\tilde{X}_i)\tilde{x}(t_{k+1})\nonumber \\
&<0\,, 
\end{align}
where the inequality comes from considering the switching law condition \eqref{eq:strat2_cond2}.

Thus, from \eqref{eq:strat2_step1}, \eqref{eq:strat2_step2} and \eqref{eq:strat2_deltaV} one has that $\dot{V}(\tilde{x},t) \leq -x(t)'C'Cx(t) +\varepsilon$ for almost all $t \in [t_k, t_{k+1}]$ and $\Delta V(\tilde{x}(t),t) <0$ at switching instants. Hence, iterating the above steps leads to $\dot{V}(\tilde{x},t) \leq -x(t)'C'Cx(t)+\varepsilon$ for almost all $t\geq t_0$ thus guaranteeing practical stability for the state $x$. Finally, integrating both sides, it holds that
\begin{equation}\label{eq:affine_cost}
    J(x_0,t) = \int_{t_0}^{t} z(\tau)'z(\tau) \mathrm{d}\tau \leq V(x(t_0),t_0) + \varepsilon(t-t_0),
\end{equation}
thus proving the statement.
\end{IEEEproof}
It is worth highlighting that, in the case of linear switched systems, i.e., $b_i=0$ for all $i\in \Omega$, the conditions of Theorem \ref{thm:main} are equivalent to those presented in Theorem \ref{thm:differential_P} with $\varepsilon$ being arbitrarily small.

Noticeably, similarly to Theorem \ref{thm:differential_P}, matrices $\tilde{P}_i(t)$ are time-varying. In fact, such matrices evolve according to the differential equation \eqref{eq:strat2_Pdot_1} in the time interval $t\in[t_k,t_k+T)$, then they achieve $\tilde{P}_i(t)=\tilde{X}_i$ at $t=t_k+T$ and remain constant until the next switching instant. Therefore, while $\tilde{P}_i(t)$ has the block-diagonal form \eqref{eq:tildeX_j} in the interval $[t_k+T,t_{k+1})$, the same structure is not necessarily guaranteed, nor required, in the interval $[t_k,t_k+T)$.
\begin{remark}[Initialization of the switching signal]
    Following the approach adopted for the switched linear case, the switching signal can be initialized as $\sigma(t_0) = \argmin_{\sigma} V(x(t_0),t_0)$ to minimize the first term of the cost \eqref{eq:affine_cost}.  \hfill$\triangledown$
\end{remark}
\begin{remark}[Average cost ultimate bound]
    From \eqref{eq:affine_cost} it is clear that, in the long run
    \begin{equation*}
        \lim_{t\to\infty} \frac{1}{t-t_0}\int_{t_0}^t z(\tau)'z(\tau) \mathrm{d}\tau \leq \varepsilon\,.
    \end{equation*}\hfill$\triangledown$
\end{remark}

The Lyapunov-Metzler condition \eqref{eq:strat2_LyapMetz} can be explicitly expressed as
    \begin{multline}
    \begin{bmatrix}
        A_i'X_i &\!\!\! 0_n \\
        b_i'X_i &\!\!\! 0
    \end{bmatrix}+
    \begin{bmatrix}
        X_i A_i &\!\!\! X_i b_i \\
        0_n' &\!\!\! 0
    \end{bmatrix}\\+
    \sum_{j\in\Omega_i} \pi_{i,j} \left( 
    \begin{bmatrix}
        e^{A_j'T}X_je^{A_jT} &\!\!\! e^{A_jT}X_jm_j \\
        m_j'X_j e^{A_jT} &\!\!\!  m_j'X_jm_j + 1
    \end{bmatrix}\right.\\+ 
    \left. \int_0^T \begin{bmatrix}
        e^{A'\tau}C'Ce^{A\tau} &\!\!\! e^{A'\tau}C'C m_j \\
        m_j'C'Ce^{A\tau} &\!\!\! m_j'C'Cm_j
    \end{bmatrix}\mathrm{d}\tau -
    \begin{bmatrix}
        X_i &\!\!\! 0_n \\
        0_n' &\!\!\! 1
    \end{bmatrix}
    \right)\\ +
    \begin{bmatrix}
        C'C &\!\!\! 0_n \\
        0_n' &\!\!\! 0
    \end{bmatrix} <\varepsilon \tilde{I}, \label{eq:extended_LM}        
    \end{multline}
with 
\begin{equation*}
    m_j \coloneqq \int_{t}^{t+T} \ee^{A_j(t+T-\tau)}b_j \mathrm{d}\tau =\int_{0}^{T} \ee^{A_j(T-\tau)}b_j \mathrm{d}\tau.
\end{equation*}
Therefore, condition \eqref{eq:strat2_LyapMetz} is equivalent to finding symmetric positive definite matrices $X_i\in \R^{n \times n}$ such that the extended Lyapunov-Metzler inequality \eqref{eq:extended_LM} is satisfied.

In the following, we provide further intuition regarding the design of the switching rule \eqref{eq:strat2}.
\begin{remark}[Interpretation of the switching law]\label{rem:strat2_logic}
    The switching law presented in Theorem \ref{thm:main} relies on the comparison of the Lyapunov function of the active subsystem at the current time with the forecast of the Lyapunov functions of the remaining subsystems $T$ time instants forward in time plus the \emph{cost-to-go} in the interval $[t,t+T]$. In fact, the switching law \eqref{eq:strat2} can be rewritten as
    \begin{subequations}
        \begin{align}
                &\sigma(x(t),t) = i \enskip \forall t\in [t_k,t_k+T ]\\
                &\sigma(x(t),t) = i \enskip \forall t>t_k+T,\,  \label{eq:V_b} \\
                &\hspace{0.3cm}\text{if}\; V_j(x(t\!+\!T),t\!+\!T)\!+\!J_j(t,t\!+\!T)\geq V_i(x(t),t),\,\forall\,j\in\Omega_i\nonumber \\
                &\sigma(x(t_{k+1}),t_{k+1})\!=\!\argmin_{j\in\Omega_i}\! V_j(x(t_{k+1}\!+\!T)\!)\!+\!J_j(t,t\!+\!T),  \label{eq:V_c}
        \end{align}
    \end{subequations}
    where $V_j(x(t),t)\coloneqq \tilde{x}'\tilde{X}_j\tilde{x}$, 
    \begin{align*}
        J_j(t,t+T)&= \int_t^{t+T} x'(\tau)C'Cx(\tau) \mathrm{d}\tau \\
        & =\int_0^{T} \tilde{x}'(t+\tau)\tilde{C}'\tilde{C}\tilde{x}(t+\tau) \mathrm{d}\tau =\tilde{x}'(t)\tilde{Y}_{2,j} \tilde{x}(t) , 
    \end{align*} 
    that is the cost over the $j$th subsystem. 
    Note that, the only way to evaluate conditions \eqref{eq:V_b} and \eqref{eq:V_c} is through the explicit computation of the Lyapunov functions and cost-to-go as presented in Theorem \ref{thm:main}. \hfill$\triangledown$
\end{remark}
 
\section{Parameter tuning and limit cases}\label{sec:tuning}
The control strategies presented in the previous sections require tuning some parameters such as the Metzler matrix $\Pi$ and the scalar parameters $\epsilon$ and $\varepsilon$ in Corollary \ref{corol:Pdot_extension} and Theorem \ref{thm:main}, respectively. The choice of these parameters must aim at minimizing the guaranteed cost upper-bound. Hence, for Theorem \ref{thm:differential_P}, the optimal choice of matrices $\Pi$ and $X_i$ is the one obtained by solving the problem
\begin{equation*}
    \min_{\Pi \in \mathcal{M}, X_1>0,\dots,X_M >0} \left\lbrace  x(t_0)'P_{\sigma(t_0)}(t_0)x(t_0) \enskip \colon \enskip \eqref{eq:strat1_LyapMetz} \right\rbrace,
\end{equation*}
which is equivalent to
\begin{multline}\label{eq:min_linear}
        \min_{\Pi \in \mathcal{M}, X_1>0,\dots,X_M >0} \biggl\{ x(t_0)'\biggl( e^{A_{\sigma(t_0)}' T}X_{\sigma(t_0)}e^{A_{\sigma(t_0)} T} \biggr. \biggr. \\
        \biggl. \biggl. + \int_{0}^{T} e^{A_{\sigma(t_0)}' \tau}C' C e^{A_{\sigma(t_0)} \tau} \mathrm{d}\tau \biggr) x(t_0)  \enskip \colon \enskip \eqref{eq:strat1_LyapMetz} \biggr\},
\end{multline}
where $\sigma(t_0)$ is selected according to Remark \ref{rem:initialization_linear}.

In the case of affine systems addressed in Theorem \ref{thm:main}, it is desirable to minimize the \emph{persistent} cost provided by the term $\varepsilon$, that is
\begin{equation}\label{eq:min_affine}
    \min_{\Pi \in \mathcal{M}, X_1>0,\dots,X_M >0} \left\lbrace\varepsilon  \enskip \colon \enskip \eqref{eq:strat2_LyapMetz} \right\rbrace.
\end{equation}
Nevertheless, determining a numerical solution for the Lyapunov–Metzler inequalities concerning the variables $\left(\Pi, \{X_1, \ldots, X_M\}\right)$ that solve problems \eqref{eq:min_linear} and \eqref{eq:min_affine} presents significant challenges and warrants further investigation. The primary difficulty arises from the nonconvex nature of the problem, attributed to the products of variables, which renders LMI solvers ineffective. 

\subsection{Simplified solution to Lyapunov-Metzler inequalities}
A potential strategy for further exploration is leveraging the specific structure of the inequalities, where \(\pi_{i,j}\) are scalars, to develop an interactive method based on relaxation. Nevertheless, as shown in \cite{LM:GeromelColaneri:2006}, a simpler, albeit more conservative, stability condition can be formulated using LMIs, making it solvable with existing techniques. The following theorem demonstrates that, by adapting Theorem \ref{thm:differential_P} to the limit case with $T=0$, and by focusing on a subclass of Metzler matrices, defined by identical diagonal elements, this objective can be achieved.
\begin{theorem}\label{thm:T0}
    Consider the switched linear system \eqref{eq:linear_sys} and assume that there exist symmetric positive definite matrices $X_i \in \R^{n\times n}$ and $\gamma >0$ so that
    \begin{equation}\label{eq:T0_LyapMetz}
        A_i' X_i + X_i A_i + \gamma (X_j - X_i) + C' C < 0, 
    \end{equation}
    for any $i \in \Omega$. Then, considering $\sigma(t^-)=i$, the switching strategy
    \begin{align}\label{eq:T0_cond}
        \sigma(t)  =  \argmin_{j\in\Omega_i}x(t)'X_jx(t) 
    \end{align}
     guarantees that the origin of system \eqref{eq:linear_sys} is globally exponentially stable. Furthermore, the performance index \eqref{eq:cost} is bounded as 
    \begin{equation}\label{eq:T0_cost}
         J(x_0,t) \leq \sum_{i\in\Omega}x(t_0)'X_{i}(t_0)x(t_0).
    \end{equation}
\end{theorem}
\begin{IEEEproof}
    The proof follows straightforwardly from \cite[Theorem 4]{LM:GeromelColaneri:2006}, and it is thus omitted.
\end{IEEEproof}
Theorem \ref{thm:T0} shows that the switching law presented in Theorem \ref{thm:differential_P} is a generalization of \cite[Theorem 4]{LM:GeromelColaneri:2006}. In fact, considering $T=0$, condition \eqref{eq:strat1_Pdot} disappears and considering $\pi_{i,i}=\gamma$ for all $i \in \Omega$, condition \eqref{eq:strat1_LyapMetz} reduces to \eqref{eq:T0_LyapMetz}. Finally, the first two terms of the switching law \eqref{eq:strat1} are discarded for $T=0$ while the third term reduces to \eqref{eq:T0_cond}. 

As suggested in \cite{LM:GeromelColaneri:2006}, the minimization problem \eqref{eq:min_affine} simplifies to 
\begin{equation}
    \min_{\gamma>0, X_1>0,\dots,X_M >0} \left\lbrace \sum_{i\in\Omega} x_0'X_ix_0 \enskip \colon \enskip \eqref{eq:T0_LyapMetz} \right\rbrace,
\end{equation}
which can be solved by using LMI and line search.

Remark \ref{rem:applicability} highlights that the results from Theorem \ref{thm:differential_P} do not require the matrices $\{A_1,\dots,A_M\}$ to be Hurwitz. Nevertheless, in the case each subsystem is asymptotically stable, the Lyapunov-Metzler condition \eqref{eq:strat1_LyapMetz} also holds for the choice of $\Pi=0$, and the strategy proposed preserves stability. The same reasoning also applies for condition \eqref{eq:strat2_LyapMetz} in Theorem \ref{thm:main}. In the case of a linear system \eqref{eq:linear_sys} characterized by a set of matrices $\{A_1,\dots,A_M\}$ being quadratically stable, it is well known that the Lyapunov-Metzler inequalities \eqref{eq:strat1_LyapMetz} admit a solution $X_1=\dots=X_M=X$ for which any switching law $\sigma(t) = i\in \Omega$ asymptotically stabilizes the switched linear system. Hence, Theorem \ref{thm:differential_P} contains, as a particular case, the quadratic stability condition.

\subsection{Existence of a Hurwitz average system}
A further interesting result can also be obtained when the switched affine system possesses a convex Hurwitz linear combination and the dwell time characterizing the switching law is close to being null. To describe such limit case, let us formalize the following property (see \cite{arbitrarySwitching:Bolzern:2004,affine:AlbeaZaccarian:2019,switched:LiberzonMorse:1999}).
\begin{property}\label{propr:average_system}
Given the simplex $\Lambda\coloneqq \{\lambda\in[0,1]^M\mid \sum_{i=1}^M\lambda_i=1\}$, there exists $\lambda\in\Lambda$, such that
\begin{equation}\label{eq:linear_combination}
b_\lambda \coloneqq\sum_{i\in\Omega}\lambda_ib_i = 0,\enskip\text{and}\enskip \; A_{\lambda}\coloneqq \sum_{i\in\Omega}\lambda_iA_i\enskip\text{is Hurwitz}
\end{equation}
 \end{property}
Note that, Property \ref{propr:average_system}, despite being a rather common requirement, is not a necessary condition for stabilization of switched affine systems, see \cite[Sec. 3.4.2]{switching:Liberzon:2003}. 

Finally, to extend the investigation to the case of switched affine systems, let
\begin{equation*}
    \tilde{A}_\lambda = \begin{bmatrix}
        A_\lambda & b_\lambda \\
        0_n' & 0
    \end{bmatrix}.
\end{equation*}
\begin{proposition}
    Consider the switched affine system \eqref{eq:affine_sys}, assume that Property \ref{propr:average_system} holds and define $\pi_{i,j}=\tfrac{1}{T} \bar{\pi}_{i,j}$, with $\bar{\pi}_{i,j}$ being fixed entries of a matrix $\bar{\Pi} \in \mathcal{M}$. Assume that there exist positive definite matrices $\tilde{X}_i$ and positive scalar $\varepsilon$ such that the following Lyapunov-Metzer equalities hold
    \begin{equation}\label{eq:lyapMetz2}
        \tilde{A}_i' \tilde{X}_i + \tilde{X}_i \tilde{A}_i + \frac{1}{T} \sum\limits_{j \in \Omega_i} \bar{\pi}_{i,j}(\tilde{Y}_{1,j}+\tilde{Y}_{2,j} - \tilde{X}_i) + \tilde{C}' \tilde{C} = \varepsilon \tilde{I},
    \end{equation}
    for any $i \in \Omega$. Then, as $T$ tends to $0$ it follows that $\tilde{X}_j=\tilde{X}_i=\tilde{X}$ for all $i,j\in\Omega$ with $\tilde{X}$ satisfying the Lyapunov equality
        $$
        \tilde{A}_\lambda'\tilde{X}+\tilde{X} \tilde{A}_\lambda+\tilde{C}'\tilde{C}=0.
        $$
        Then, the switching law \eqref{eq:strat2} makes the origin of system \eqref{eq:affine_sys} a globally asymptotically stable switched equilibrium. 

\end{proposition}	
\begin{IEEEproof}
    Let us consider the equality \eqref{eq:lyapMetz2}. Multiplying both sides by $T$, for $T \to 0$, it holds
    \begin{equation}
        \sum_{j\in \Omega_i}\bar\pi_{i,j}(\tilde{X}_j-\tilde{X}_i)= \sum_{j\in \Omega}\bar\pi_{i,j}\tilde{X}_j=0_{n+1\times n+1} \quad \forall i \in \Omega,
    \end{equation}
    where the first equality derives from $\bar{\Pi} \in \mathcal{M}$.
    Letting $\zeta\in\R^{n+1}$, one can write
    \begin{equation}
        \sum_{j\in \Omega}\bar\pi_{i,j}\zeta'\tilde{X}_j\zeta=\sum_{j\in \Omega}\bar\pi_{i,j}\eta_j=0 \quad \forall i \in \Omega,
    \end{equation}
    which in turn implies that $\bar{\Pi}\eta=0$, with 
    $\bar{\Pi}\in\mathcal{M}$, and $\eta\in\mathbb{R}^M$. 
    By virtue of the Frobenius-Perron theorem, the eigenvector associated with the null eigenvalue of $\bar{\Pi}$ is strictly positive, leading to the conclusion that there always exists $\eta=\bar{\eta}{1}_M$, $\bar {\eta}>0$, so that $\zeta'(\tilde{X}_i-\tilde{X}_j)\zeta=0$, $\forall\,\zeta\in\R^{n+1}$ for $i,j\in\Omega$. This means that $\tilde{X}_j=\tilde{X}_i=\tilde{X}$ \cite{LM:GeromelColaneri:2006}. 
    
    Note that, for $T\to 0$, we can set $\varepsilon=0$. Now, multiplying the Lyapunov-Metlzer equations \eqref{eq:lyapMetz2} by the elements of the left Frobenius eigenvector $\lambda_i$, with $\lambda \in \Lambda$, substituting matrices $\tilde{X}_i$ with $\tilde{X}$ and summing over $i\in\Omega$, one has
    $$
    \sum_{i\in\Omega} \lambda_i\left( \tilde{A}_i'\tilde{X} +\tilde{X} \tilde{A}_i+\tilde{C}'\tilde{C}\right)=0 
    $$
    which implies
    \begin{align}
        \left(\sum_{i\in\Omega} \lambda_{i} \tilde{A}_i'\right)\tilde{X} +\tilde{X} \left(\sum_{i\in\Omega}\lambda_{i} \tilde{A}_i\right)+\tilde{C}'\tilde{C}&= \nonumber \\
        \tilde{A}_\lambda'\tilde{X} +\tilde{X} \tilde{A}_\lambda+\tilde{C}'\tilde{C}&=0
    \end{align}
which, according to Property 1, in turn implies
$$
A_\lambda'X +X A_\lambda+C'C=0,
$$
where $X$ is adopted to construct $\tilde{X}$ as in \eqref{eq:tildeX_j}.

Hence, taking the switching rule \eqref{eq:strat2}, by subtracting $\tilde{x}(t_{k})'\tilde{X}_i \tilde{x}(t_{k})$ to the argument of \eqref{eq:strat2_cond3} and dividing by $T$, for $T\to 0$, it reduces to 
$$
\sigma(t_k) = \argmin_{j\in\Omega}\dot{v}(x(t),t),
$$
where $v(x(t),t)=x'(t)X x(t)$.
By similar reasoning as in \cite[Theorem 3]{LM:GeromelColaneri:2006}, and exploiting the assumption that $b_{\lambda}=0$, we have that 
    \[
    \dot{v }(x(t),t)\leq -x'(t)C'Cx(t).
    \]
Since the Lyapunov function $v(x(t),t)$ is radially unbounded, then the equilibrium $x=0$ of \eqref{eq:linear_sys} is globally asymptotically stable.
\end{IEEEproof}

\subsection{Existence and design of Metzler matrix}
Finally, a relevant point to be discussed now concerns the existence of a solution of the Lyapunov–Metzler inequalities \eqref{eq:strat2_LyapMetz} with respect to the variables $\Pi$ and $\{\tilde{X}_1,\dots,\tilde{X}_m\}$. It is not difficult to show through standard Kronecker calculus that, for fixed $\Pi \in \mathcal{M}$, a solution with respect to the remaining variables exists if and only if the $m\cdot (n+1)^2$-dimensional square matrix $\mathcal{J} = \mathcal{A}+\mathcal{B}(\Pi)$ is Hurwitz, with
\begin{equation*}
    \mathcal{A} = \begin{bmatrix}
        \tilde{A}_1'\oplus\tilde{A}_1' & 0 & \dots & 0 \\
        0 & \tilde{A}_2'\oplus\tilde{A}_2' & \dots & 0 \\
        0 & 0 & \ddots & 0 \\
        0 & 0 & \dots & \tilde{A}_m'\oplus\tilde{A}_m'
    \end{bmatrix},
\end{equation*}
and
\begin{equation*}
    \mathcal{B}(\Pi) = \Pi_d \otimes I_{(n+1)^2} + \left[\left(\Pi - \Pi_d\right)\otimes I_{(n+1)^2}\right]e^{\mathcal{A}T},
\end{equation*}
with $\Pi_d$ being the diagonal matrix with elements being those of the diagonal of $\Pi$. Hence, the existence of a solution to \eqref{eq:strat2_LyapMetz} reduces to the existence of $\Pi\in\mathcal{M}$ making matrix $\mathcal{J}$ asymptotically stable. A similar existence condition holds for the Lyapunov-Metzer inequality \eqref{eq:strat1_LyapMetz} where terms $\tilde{A}_i$ in $\mathcal{A}$, $i\in \Omega$, are replaced with terms $A_i$.

A possible approach to turn the optimization problems \eqref{eq:min_linear} and \eqref{eq:min_affine} into LMI is to fix matrix $\Pi$ and solve for matrices $X_i$. With regards to the choice of the matrix $\Pi$ in Theorem \ref{thm:differential_P}, Corollary \ref{corol:Pdot_extension} and Theorem \ref{thm:main}, different approaches can be pursued, depending on the nature of the switched system. In general, two distinct families of systems can be identified: those for which Property \ref{propr:average_system} is satisfied, and those for which there does not exist such convex combination satisfying condition \eqref{eq:linear_combination}. In the former case, indicating with $\lambda$ the vector that guarantees satisfaction of Property \ref{propr:average_system}, matrix $\Pi\in \mathcal{M}$ can be chosen so that $\lambda'\Pi=0$. Otherwise, if there does not exist any convex combination that satisfies Property \ref{propr:average_system}, one can rely on the fact that the matrix $\Pi$ can be interpreted as the probability rate matrix of the Markov process associated with the switching sequence. This in turn implies that the elements $\pi_{i,j}$ of $\Pi$ are selected as the expected probability rate of switching to the $j$th subsystem when the $i$th subsystem is active, i.e., $\pi_{i,j} = \mathrm{Prob}[\sigma(t+ \mathrm{d}t) = j \,|\, \sigma(t) = i]$, for $i\neq j$ so that 
\begin{equation*}
    T_i = \frac{1}{\sum_{j\neq i}\pi_{i,j}},
\end{equation*}
is the average dwell time for the $i$th mode \cite{markov:Bolzern:2010}. See, for instance, the congestion control problem in the next section. Noticeably, the latter approach requires the knowledge of the expected desirable switching sequence.

\section{Illustrative Examples}\label{sec:examples}
    The switching strategies detailed in Section \ref{sec:switching_linear} and Section \ref{sec:switching_affine} are evaluated through various examples. The strategy formulated for switched linear systems, as outlined in Theorem \ref{thm:differential_P}, is applied to stabilize a switched system comprising two unstable linear subsystems. Subsequently, the switching strategy from Theorem \ref{thm:main} is implemented for the control of a boost-boost converter and to address a traffic congestion problem. Notably, while a convex combination of the subsystems exists for the boost-boost converter such that condition \eqref{eq:linear_combination} is satisfied, no such convex combination exists for the traffic congestion scenario. Nevertheless, the proposed switching strategy manages to orchestrate the congestion scenario properly.
    \subsection{Unstable subsystems}
\begin{figure}[htb]
        \centering
        \includegraphics[width=0.9\columnwidth]{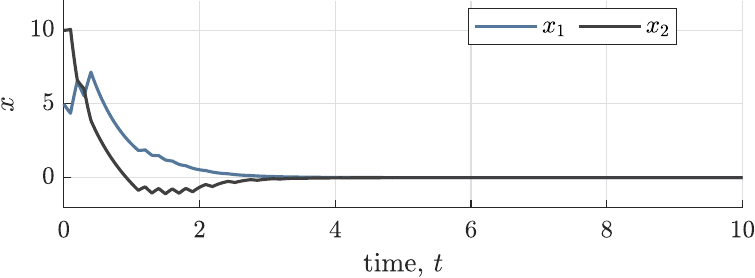}
        \caption{Time-evolution of the state $(x_1(t),x_2(t))$ for the switched linear system.}
        \label{fig:unstable_linear}
    \end{figure}

    Let us consider the switched linear system \eqref{eq:linear_sys}, with
    \begin{equation*}
        A_1 = \begin{bmatrix}
            -2 & 0.3 \\
            -2 & 1
        \end{bmatrix}, \enskip A_2 = \begin{bmatrix}
            1 & 2 \\
            -0.3 & -4
        \end{bmatrix}, \enskip C = \begin{bmatrix}
            1 & 0 \\
            0 & 1
        \end{bmatrix}.
    \end{equation*}
    Both systems are unstable, however, it is easy to verify that the convex combination obtained with $\lambda_1=\lambda_2=0.5$ generates a stable system. Then, matrix $\Pi$ can be chosen of the form
    \begin{equation}\label{eq:Pi_unstable_example}
        \Pi = \alpha \begin{bmatrix}
            -1 & 1 \\
            1 & -1
        \end{bmatrix}, \quad \alpha >0.
    \end{equation}
    The simulation results, derived using the switching law in Theorem \ref{thm:differential_P}, with dwell time $T=0.1$ and the initial condition $x_0 = [5\,\, 10]'$, are shown in Fig. \ref{fig:unstable_linear}. The figure demonstrates that the state trajectories converge asymptotically to the origin of the state space, despite the inherent instability of the two subsystems.

\begin{figure}[htb]
        \centering
        \includegraphics[width=0.9\columnwidth]{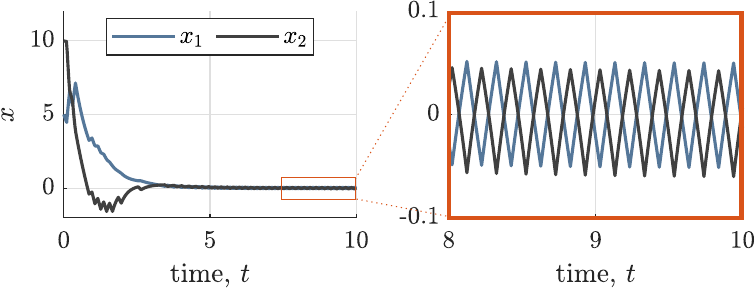}
        \caption{Time-evolution of the state $(x_1(t),x_2(t))$ for the affine switched system.}
        \label{fig:unstable_affine}
    \end{figure}

\begin{figure}[htb]
        \centering
        \includegraphics[width=0.9\columnwidth]{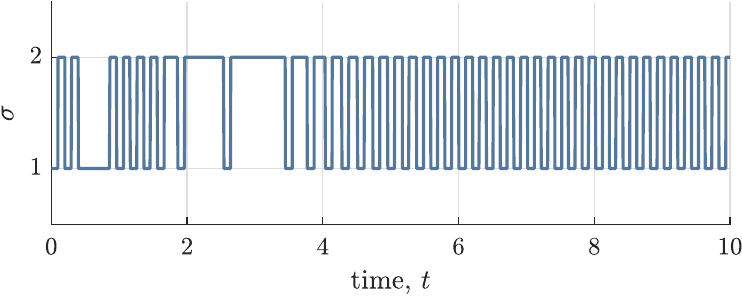}
        \caption{Time-evolution of the switching signal $\sigma(t)$ for the affine switched system.}
        \label{fig:unstable_affine_sigma}
    \end{figure}

    This numerical example can be extended by considering the switched affine system \eqref{eq:affine_sys} with
    \begin{equation*}
        b_1 = \begin{bmatrix}
            1 \\
            -1
        \end{bmatrix}, \quad
        b_2 = \begin{bmatrix}
            -1 \\
            1
        \end{bmatrix}.
    \end{equation*}
    It is easy to see that the same convex combination as in the linear case satisfies condition \eqref{eq:linear_combination}, hence the matrix $\Pi$ can be designed again as in \eqref{eq:Pi_unstable_example}. In this scenario, by applying the switching law from Theorem \ref{thm:main}, with dwell time $T=0.1$ and the initial condition $x_0 = [5\,\, 10]'$, the state trajectories converge towards a neighborhood of the origin and oscillate around it, as shown in Fig. \ref{fig:unstable_affine}. The switching signal alternates between $\sigma(t)=1$ and $\sigma(t)=2$ ultimately converging to a periodic solution (see Fig. \ref{fig:unstable_affine_sigma}).

	\subsection{Boost-boost converter}
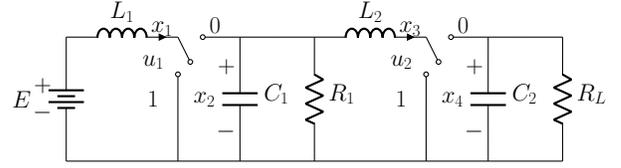
\begin{figure}[!t]
	\centering
	\ctikzset{bipoles/length=4.75cm, font=\fontsize{55}{0}\selectfont}
	\begin{circuitikz} [scale=0.165, transform shape]
		\draw (0,0) 	to[short] (40,0);
		\draw (0,10) 	to[battery, v_=$E$, voltage/american label distance=1.8] (0,0);
		\draw (0,10) 	to[L=$L_1$,i=$x_1$] (9,10);
		\draw (9,10) 	to[short,-o](10,8);
            \draw (9,0) 	to[short,-o](9,7);
		\draw (14,10) 	to[short,-o](11,10);
		\node at (7,8) (u1){$u_1$};
		\node at (7,5) (u11){$1$};
		\node at (12,11) (u10){$0$};
		\draw (14,10)	to[C=$C_{1}$,v_=$x_2$] (14,0);
		\draw (20,10)	to[R=$R_1$] (20,0);
		\draw (14,10) 	to[short] (20,10);
		\draw (20,10) 	to[L=$L_2$,i=$x_3$] (29,10);
		\draw (29,10) 	to[short,-o](30,8);
		\draw (29,0) 	to[short,-o](29,7);
		\draw (34,10) 	to[short,-o](31,10);
		\node at (27,8) (u2){$u_2$};
		\node at (27,5) (u21){$1$};
		\node at (32,11) (u20){$0$};
		\draw (34,10)	to[C=$C_{2}$,v_=$x_4$] (34,0);
		\draw (40,10)	to[R=$R_L$] (40,0);
		\draw (34,10) 	to[short] (40,10);
	\end{circuitikz}
	\caption{Boost-boost converter topology.}
	\label{fig:boostboost_scheme}
\end{figure}

    The switching strategy proposed in Theorem \ref{thm:main} is particularly well-suited for controlling DC-DC switched power converters. These converters naturally possess a switched structure, and the limited commutation frequency of their switching devices necessitates the design of a discontinuous control law that incorporates dwell time. Additionally, most DC-DC switched power converters can be modeled as switched affine systems. In this section, we focus on the control of the boost-boost converter, as discussed in \cite[§2.10]{power_conv:SiraRamirez:2006}. The boost-boost converter features two switching elements (see Fig. \ref{fig:boostboost_scheme}), resulting in four distinct configurations of the converter.

\begin{figure}[htb]
        \centering
        \includegraphics[width=0.9\columnwidth]{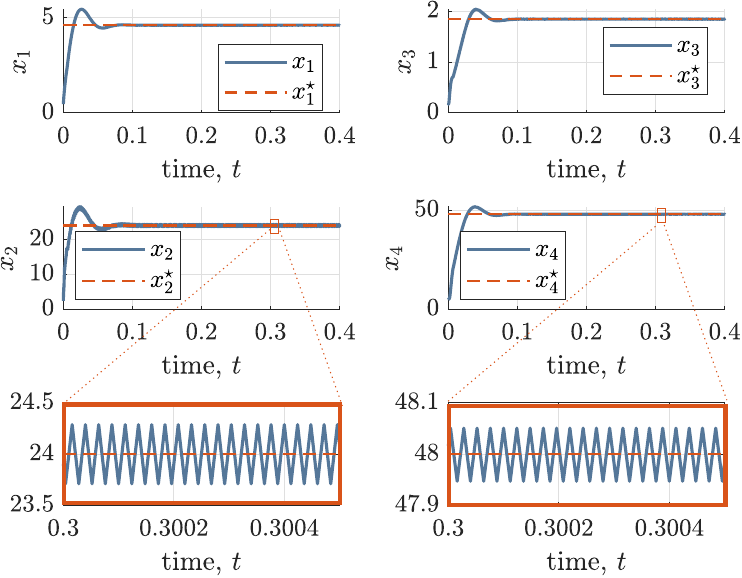}
        \caption{Time-evolution of the boost-boost converter states. Top left: current $x_1(t)$. Top right: first stage voltage $x_2(t)$. Bottom left: current $x_3(t)$. Bottom right: second stage voltage $x_4(t)$.}
        \label{fig:boost_plot}
    \end{figure}

    Therefore, such converter can be modeled as a switched affine system of the form \eqref{eq:affine_sys} with $\sigma \in \{1,2,3,4\}$ and   
    \begin{alignat*}{2}
        & A_1 = \begin{bmatrix}
            0 & -\frac{1}{L} & 0 & 0 \\
            \frac{1}{C_1} & -\frac{1}{R_1 C_1} & -\frac{1}{C_1} & 0 \\
            0 & \frac{1}{L_2} & 0 & -\frac{1}{L_2} \\
            0 & 0 & \frac{1}{C_2} & -\frac{1}{R_L C_2}
        \end{bmatrix}, \enskip &&b_1 = \begin{bmatrix}
            \frac{E}{L_1} \\
            0 \\
            0 \\
            0
        \end{bmatrix},\\
        & A_2 = \begin{bmatrix}
            0 & 0 & 0 & 0 \\
            0 & -\frac{1}{R_1 C_1} & -\frac{1}{C_1} & 0 \\
            0 & \frac{1}{L_2} & 0 & -\frac{1}{L_2} \\
            0 & 0 & \frac{1}{C_2} & -\frac{1}{R_L C_2}
        \end{bmatrix}, \enskip &&b_2=b_1, \\
        & A_3 = \begin{bmatrix}
            0 & -\frac{1}{L} & 0 & 0 \\
            \frac{1}{C_1} & -\frac{1}{R_1 C_1} & -\frac{1}{C_1} & 0 \\
            0 & \frac{1}{L_2} & 0 & 0 \\
            0 & 0 & 0 & -\frac{1}{R_L C_2}
        \end{bmatrix}, \enskip &&b_3=b_1,\\
        & A_4 = \begin{bmatrix}
            0 & 0 & 0 & 0 \\
            0 & -\frac{1}{R_1 C_1} & -\frac{1}{C_1} & 0 \\
            0 & \frac{1}{L_2} & 0 & 0 \\
            0 & 0 & 0 & -\frac{1}{R_L C_2}
        \end{bmatrix}, \enskip &&b_4=b_1,
    \end{alignat*}
    where $x_1$ is the current through the inductor $L_1$, $x_2$ is the output voltage at the first stage of conversion, i.e., the voltage across capacitor $C_1$, $x_3$ represents the current through the inductor $L_2$, while $x_4$ indicates the output voltage at the second stage of conversion, i.e., the load voltage across capacitor $C_2$, or equivalently the load voltage. 
    The switching signal $\sigma(t)$ denotes the configurations of the switch pair $(u_1,u_2)$, with $\sigma=1$ corresponding to the configuration  $(0,0)$, $\sigma=2$ to $(1,0)$, $\sigma=3$ to $(0,1)$ and, finally, and $\sigma=4$ to $(1,1)$. The objective of the boost-boost converter is to regulate the voltages at the two output stages, specifically $x_2$ and $x_4$ to their respective reference values, $x_2^\star$ and $x_4^\star$. Hence, for given $x_2^\star$ and $x_4^\star$, a parametrization of the equilibrium point in terms of the steady state output voltages is obtained as
    \begin{equation*}
        x_1^\star = \frac{1}{R_1}\frac{x_2^{\star^2}}{E} + \frac{1}{R_L}\frac{x_4^{\star^2}}{E}, \quad x_3^\star = \frac{1}{R_L}\frac{x_4^{\star^2}}{x_2^\star}.
    \end{equation*}
    Therefore, the control objective is to regulate to zero the error variable $x^e = x - x^\star$,  $x^\star = [x_1^\star\,\,x_2^\star\,\,x_3^\star\,\,x_4^\star]'$, obeying to the switched differential equation
    \begin{equation}\label{eq:boostboost_shifted}
        \dot{x}^e(t) = A_{\sigma(t)}x^e(t)+b^e_{\sigma(t)}, \quad z^e(t) = Cx^e(t)
    \end{equation}
    where $b^e_{\sigma(t)} = A_{\sigma(t)}x^\star+b_{\sigma(t)}$. The numerical values for the converter parameters are selected as specified in \cite[§2.10]{power_conv:SiraRamirez:2006}. Given the input voltage $E=12$ V, with $x_2^\star = 24$ V and $x_4^\star = 48$ V, the convex combination obtained with $\lambda_i=0.25$, $i=\{1,\dots, 4\}$ results in a stable convex combination for system \eqref{eq:boostboost_shifted}. The matrix $\Pi$ in \eqref{eq:strat2_LyapMetz} can be chosen as any matrix satisfying $ \lambda'\Pi = 0$. Due to the high switching frequency required for DC-DC power converters, the dwell time is set to $T=10^{-5}$ s, while the initial condition is chosen as $x(0) = [0.46 \, \,2.40 \, \, 0.18 \, \,4.80]'$. The performance of the proposed switching law is depicted in Fig. \ref{fig:boost_plot}, demonstrating that the states smoothly converge towards the desired reference values and oscillate around them.
    
    The performance of the proposed switching law is then compared with  the algorithms presented in \cite{affine:AlbeaZaccarian:2019} and in \cite{switched:Egidio:2020}. While the former adopts a hybrid framework formulation to enforce a dwell-time switching, the latter relies on the definition of a periodic trajectory towards which the state is attracted asymptotically. The metric considered for the evaluation of the three algorithms is the cost functional \eqref{eq:cost} evaluated for the error variable $z^e$ within the time interval $[0,\,0.4]$, i.e.,
    \begin{equation*}
        J(x_0^e,t) = \int_{0}^{0.4} z^{e\prime}(\tau)z^e(\tau) \mathrm{d}\tau = \int_{0}^{0.4} x^{e\prime}(\tau)C'Cx^e(\tau) \mathrm{d}\tau,
    \end{equation*}
    where matrix $C$ has been chosen as 
    \begin{equation*}
        C = \begin{bmatrix}
            0 & x_2^{\star^{-1}} & 0 & 0 \\
            0 & 0 & 0 & x_4^{\star^{-1}}
        \end{bmatrix},
    \end{equation*}
    to normalize the error variables with respect to the steady state values. The values obtained for the chosen metric are shown in Table \ref{tab:cost}, where it is seen that our approach achieves performance comparable to the two considered literature algorithms. 
    \begin{table}[htb]
        \centering
        \caption{Cost evaluation for our proposal, and algorithms \cite{affine:AlbeaZaccarian:2019} and \cite{switched:Egidio:2020}}
        \begin{tabular}{ccc}
        \toprule
            Theorem \ref{thm:main}  &  \cite{affine:AlbeaZaccarian:2019}  & \cite{switched:Egidio:2020}  \\
             $8.966 \cdot 10^{-3}$ & $8.971 \cdot 10^{-3}$  & $8.809 \cdot 10^{-3}$ \\
             \bottomrule
        \end{tabular}
        \label{tab:cost}
    \end{table}
    
    Nevertheless, determining the superiority of any of the three algorithms is challenging, as the results may vary depending on the specific values of their tuning parameters or the particular systems to which they are applied.
 
	\subsection{Congestion control}
\begin{figure}[htb]
        \centering
        \includegraphics[width=0.7\columnwidth]{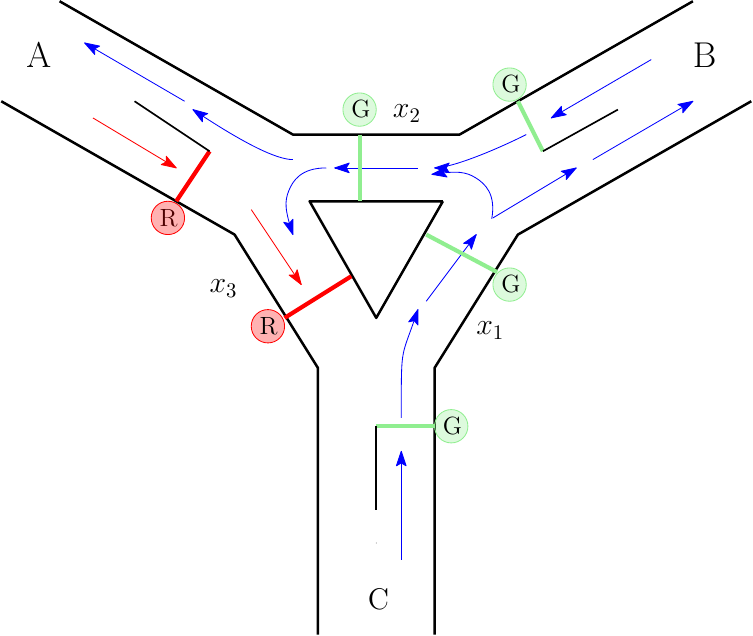}
        \caption{The traffic control problem.}
        \label{fig:congestion_scheme}
    \end{figure}

    Consider a traffic management scenario at an intersection, illustrated in Fig. \ref{fig:congestion_scheme}. Three major roads (A, B, and C) merge into a ``triangular junction" regulated by traffic signals. Three buffer variables, $x_1$, $x_2$, and $x_3$, indicate the number of vehicles waiting at the respective traffic signals within the triangular junction. The traffic signal configurations are assumed to be symmetric. In the first configuration, shown in Fig. \ref{fig:congestion_scheme}, the traffic signals corresponding to $x_1$, $x_2$, B, and C are green, while those corresponding to $x_3$ and A are red. Accordingly,
    \begin{itemize}
        \item the buffer variable $x_3$ increases proportionally (this is described by the positive constant $\beta$) with respect to $x_2$,
        \item the buffer variable $x_2$ remains approximately constant, receiving inflow from both road B and buffer $x_1$, while providing outflow to road A and buffer $x_3$,
        \item the buffer variable $x_1$  exponentially decreases as the inflow from road C is directed entirely towards $x_2$ and road B (this is due to $-\gamma<0$) . This exponential decrease accounts for the initial transient effect caused by traffic signal changes.
    \end{itemize}
    The other two configurations are obtained by a circular rotation of $x_1$, $x_2$ and $x_3$ (as well as of A, B and C).     
    The system description is completed by considering the effect of a constant input, i.e., the incoming traffic. Hence, the congestion control problem can be modeled as a switched affine system of the form \eqref{eq:affine_sys}, with     
    \begin{alignat*}{2}
        & A_1 = \begin{bmatrix}
            -\gamma & 0 & 0 \\
            0 & 0 & 0 \\
            0 & \beta & 0
        \end{bmatrix}, \enskip && b_1 = \begin{bmatrix}
            1 \\
            1 \\
            1 
        \end{bmatrix}, \\
        & A_2 = \begin{bmatrix}
            0 & 0 & \beta \\
            0 & -\gamma & 0 \\
            0 & 0 & 0
        \end{bmatrix}, \enskip && b_2 = b_1, \\
        & A_3 = \begin{bmatrix}
            0 & 0 & 0 \\
            \beta & 0 & 0 \\
            0 & 0 & -\gamma
        \end{bmatrix}, \enskip && b_3 = b_1, \\
        & C = \begin{bmatrix}
            1 & 0 & 0 \\
            0 & 1 & 0 \\
            0 & 0 & 1
        \end{bmatrix},
    \end{alignat*}
    with $\gamma = 1$ and $\beta = 1.1$. In \cite{traffic:Blanchini:2012}, it is demonstrated that there does not exist any convex combination  $A_\lambda=\lambda_1 A_1+\lambda_2 A_2 + \lambda_3 A_3$, such that $A_\lambda$ is Hurwitz, thereby failing to satisfy condition  \eqref{eq:linear_combination}. Consequently, switching strategies that depend on the existence of a Hurwitz convex combination of the matrices $A_i$ (such as \cite{arbitrarySwitching:Bolzern:2004,affine:AlbeaZaccarian:2019,timeevent:Albea:2021}) cannot be applied to the congestion control problem. A periodic (open-loop) dwell-time switching law implementing the circular activation order of the traffic lights as $3,2,1,3,2,1,\dots$ is proposed in \cite{traffic:Blanchini:2012}, and it is proved that this switching law stabilizes the switched affine system for dwell-time $T>0.19$. This commutation sequence implies that the ``red light" is imposed according to the circular order $3,2,1,3,2,1,\dots$ . The same study shows that alternative switching sequences not only degrade system performance but can also render it unstable. By leveraging the insights provided in Section \ref{sec:tuning}, the matrix $\Pi$ is set according to the expected probability of switching from one operating mode to the next one. Hence, for the congestion control problem, $\Pi$ is set as
    \begin{equation*}
        \Pi = \alpha\begin{bmatrix}
            -1 & 0 & 1 \\
            1 & -1 & 0 \\
            0 & 1 & -1
        \end{bmatrix}, \quad \alpha >0,
    \end{equation*}
    with each row of $\Pi$ representing the probability of the Markovian chain associated to the expected sequence $3,2,1,3,2,1,\dots$. 

\begin{figure}[htb]
        \centering
        \includegraphics[width=0.9\columnwidth]{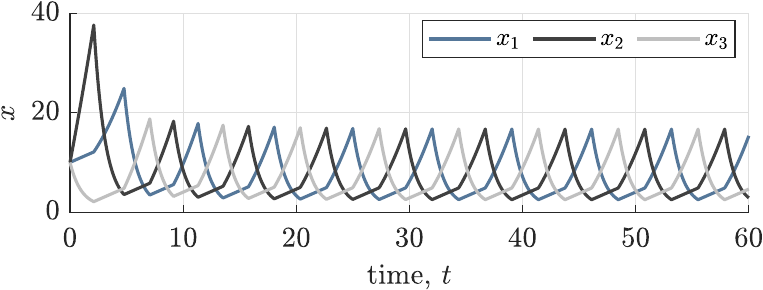}
        \caption{Number of vehicles waiting at the three traffic lights inside the triangular loop.}
        \label{fig:congestion_x}
    \end{figure}

\begin{figure}[htb]
        \centering
        \includegraphics[width=0.45\columnwidth]{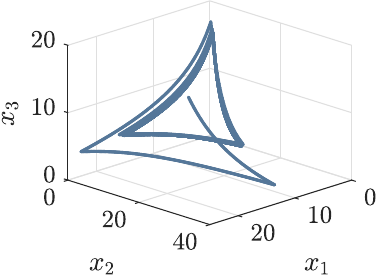}
        \caption{State trajectory in $(x_1,x_2,x_3)$ space for the congestion control problem.}
        \label{fig:congestion_3D}
    \end{figure}

    Similarly to the approach in \cite{traffic:Blanchini:2012}, the proposed switching strategy is tested in simulation with $T=2.1$ and $x(0) = [10\,\,10\,\,10]'$. The resulting state trajectories are presented in Fig. \ref{fig:congestion_x}. As illustrated, the state converges to a limit cycle, while the switching signal $\sigma(t)$, generated by the switching strategy in Theorem \ref{thm:main}, autonomously enforces the sequence $3,2,1,3,2,1,\dots$, as proposed in \cite{traffic:Blanchini:2012}. A three-dimensional representation of the state trajectories is shown in Fig. \ref{fig:congestion_3D} to further emphasize the cyclic behavior induced by the switching signal.
      
    \begin{figure}[htb]
        \centering
        \includegraphics[width=0.9\columnwidth]{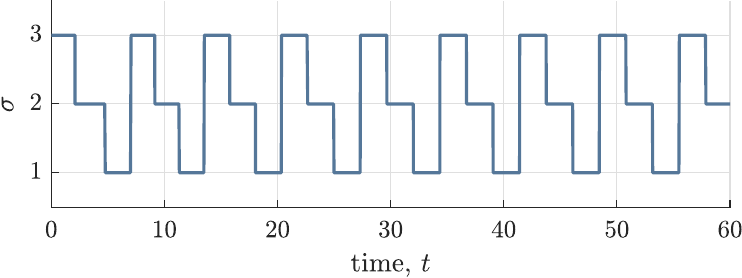}
        \caption{Switching signal $\sigma(t)$ for the congestion control problem.}
        \label{fig:congestion_sigma}
    \end{figure}
	
	\section{Conclusions}\label{sec:conclusions}
In this paper we have studied the stabilization of linear switched affine systems under dwell-time constraint about a neighborhood of the origin.
The synthesis of two state-dependent switching rules is presented. First, the case of linear switched systems has been considered to regulate their state to the origin. Then, the case of  switched affine systems has been  addressed, proving practical stabilization of the systems state in a vicinity of the origin. Both the strategies rely on the solution of differential Lyapunov inequalities and a Lyapunov-Metzler inequalities, allowing us to guarantee a bound on the cost associated with the switching control laws.
Special attention has been devoted to the minimization of such a bound in the case of affine systems, in order to guarantee suitable performance of the system response. Moreover, guidelines on the parameters tuning have been provided. 
The general validity of the proposed methods relies on their independence on the existence of a convex combination of subsystems that generates a Hurwitz average linear system.
The theoretical results have been assessed in simulation and fairly compared with other methodologies in the literature. 

An area of further work involves the application of these methods to the case of systems relying on polytopic Lyapunov functions\cite{polyhedral:Blanchini:2017}.

	\bibliographystyle{IEEEtran}
	\bibliography{biblio}

\vspace*{-1cm}
\begin{IEEEbiography}
[{\includegraphics[width=1in,height=1.25 in,clip,keepaspectratio]{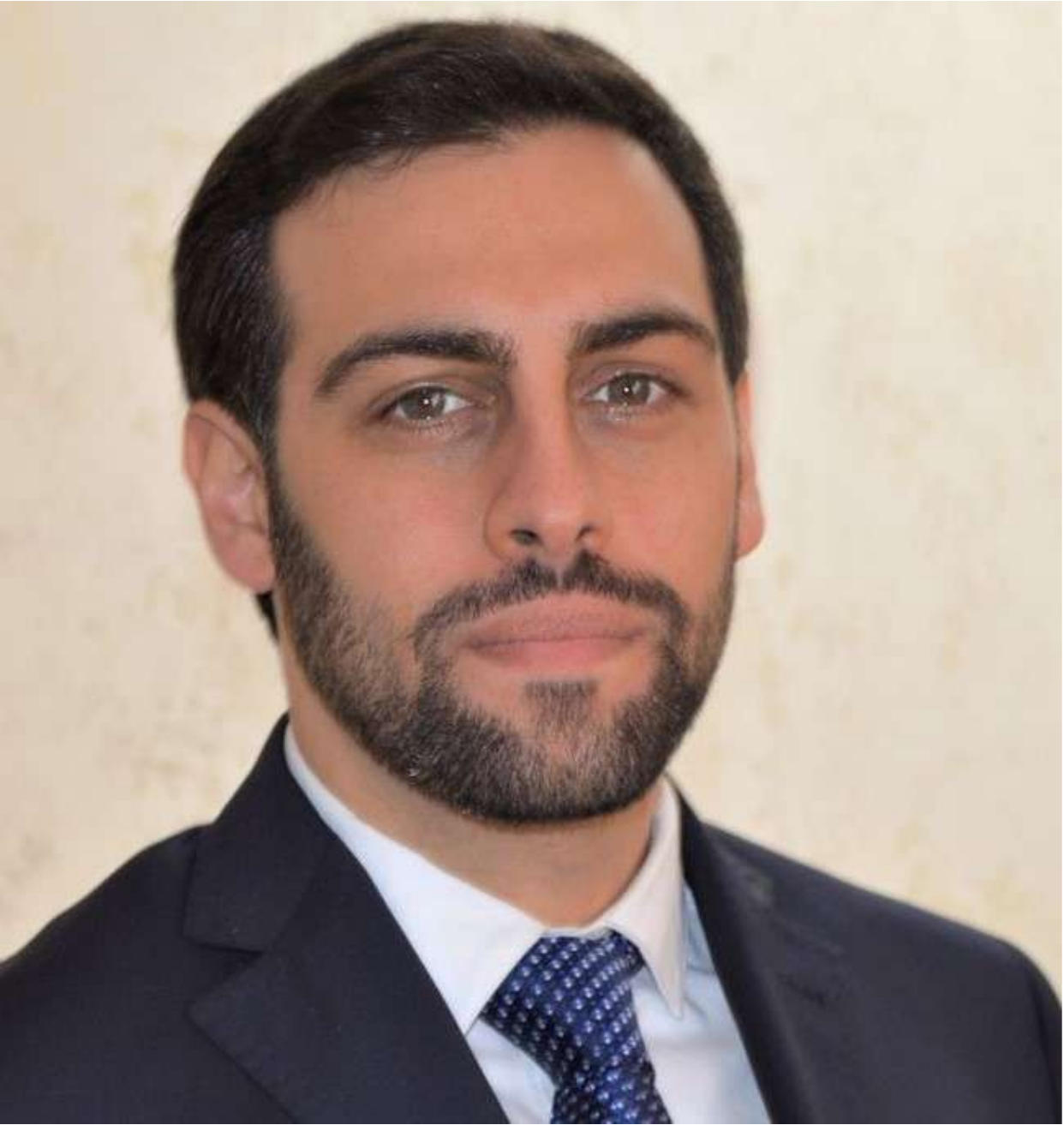}}]{Antonio Russo} (M'20) is an assistant professor of automatic control at Università degli Studi della Campania ``Luigi Vanvitelli". He received the bachelor's and master's degree's summa cum laude in Computer Science Engineering and the Ph.D. in Industrial and Information Engineering from Università degli Studi della Campania in 2015, 2017, and 2021, respectively. From January to June 2019, he was with the Coordinated Science Laboratory at the University of Illinois at Urbana-Champaign, US, and from July to October 2024 he was with the Centro Internacional Franco Argentino de Ciencias de la Información y de Sistemas, in Rosario, Argentina. Since 2022 he has been member of the conference editorial board of the European Control Association. At present, he is Associate Editor of the Franklin Open journal. His research interests include stability analysis of nonlinear switched and hybrid systems, sliding mode control and control of power systems in the framework of aircraft electrification.
\end{IEEEbiography}

\begin{IEEEbiography}
[{\includegraphics[width=1in,height=1.25 in,clip,keepaspectratio]{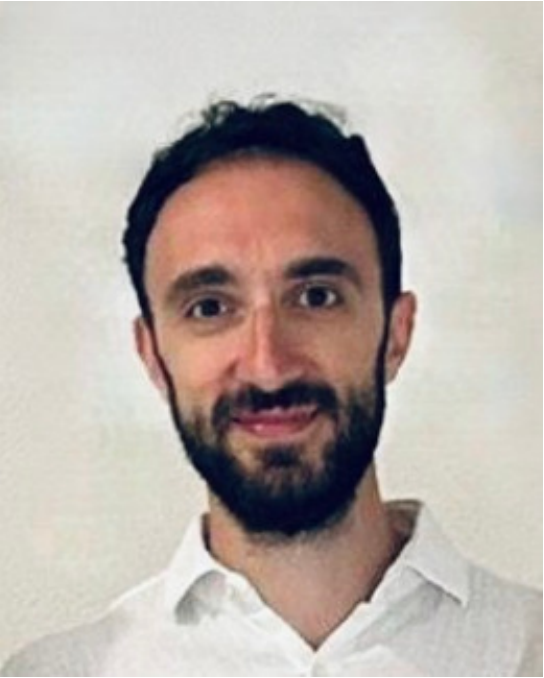}}]{Gian Paolo Incremona} (M'10, S'23) is associate professor of automatic control at Politecnico di Milano. He was a student of the Almo Collegio Borromeo of Pavia, and of the Institute for Advanced Studies IUSS of Pavia. He received the bachelor's and master's degree's summa cum laude in Electric Engineering, and the Ph.D. degree in Electronics, Electric and Computer Engineering from the University of Pavia in 2010, 2012 and 2016, respectively. From October to December 2014, he was with the Dynamics and Control Group at the Eindhoven Technology University, The Netherlands. He was a recipient of the 2018 Best Young Author Paper Award from the Italian Chapter of the IEEE Control Systems Society, and since 2018 he has been a member of the conference editorial boards of the IEEE Control System Society and of the European Control Association. At present, he is Associate Editor of the journal Nonlinear Analysis: Hybrid Systems and of International Journal of Control. His research is focused on sliding mode control, model predictive control and switched systems with application mainly to train control, robotics and power plants.
\end{IEEEbiography}

\begin{IEEEbiography}
[{\includegraphics[width=1in,height=1.25 in,clip,keepaspectratio]{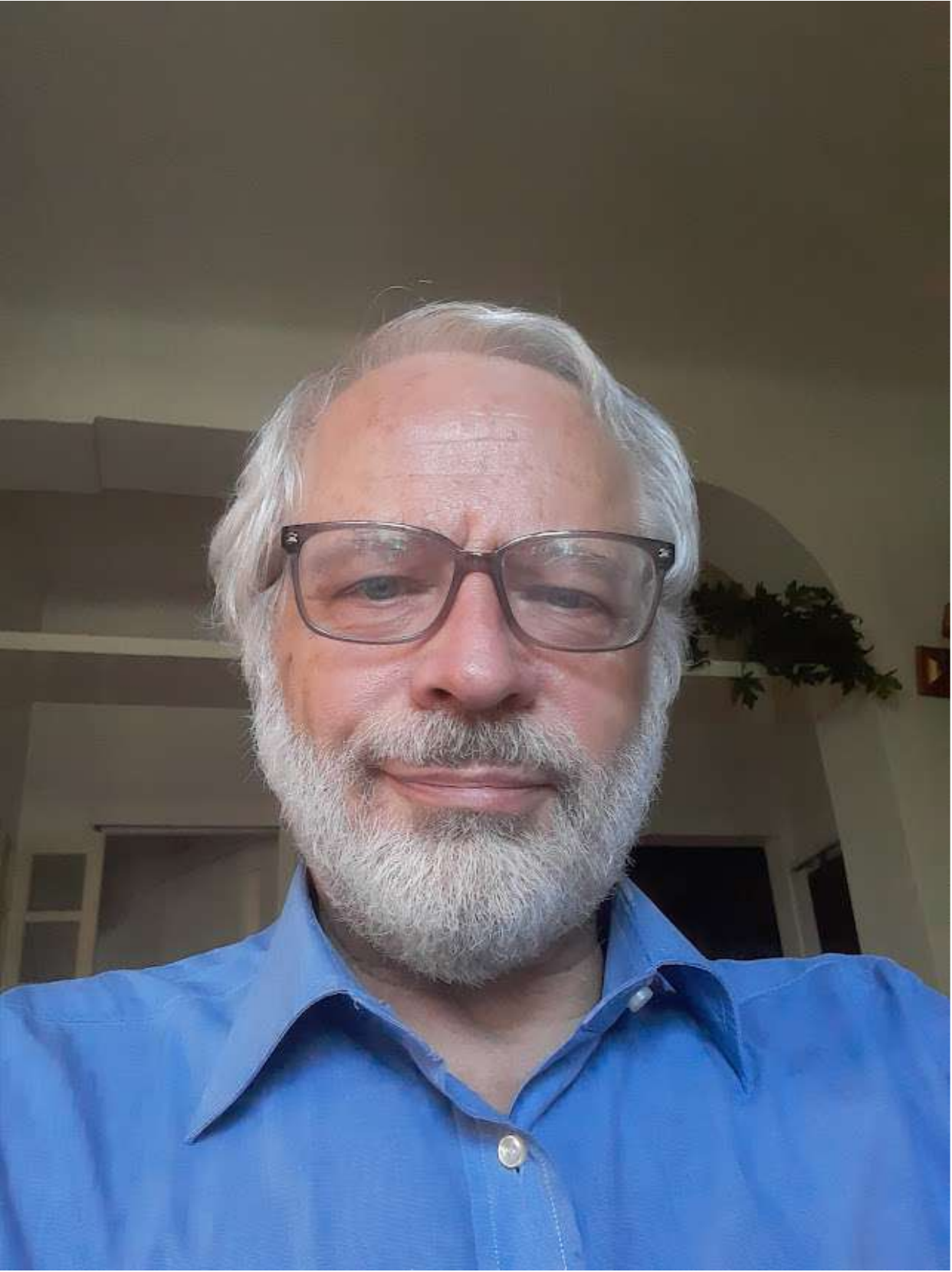}}]{Patrizio Colaneri} (Fellow, IEEE) was born in Palmoli, Italy, on October 12nd 1956. He received the Laurea degree in electrical engineering from the Politecnico di Milano, Milan, Italy, in 1981, the Ph.D. degree in automatic control from the Italian Ministry of Education and Research, Rome, Italy, in 1988. He is currently a Full Professor of automatic control with the Politecnico di Milano, where he served as the Head of the Ph.D. School on ICT from 2007 to 2009. He held visiting positions at the University of Maryland, College Park, MD, USA; the Hamilton Institute, National University of Ireland, Dublin, Ireland; and the Institute for Design and Control of Mechatronical Systems, Johannes Kepler University, Linz, Austria. His main research interests are in the area of periodic systems and control, robust filtering and control, switching control, and railway automation. He authored/coauthored seven books (three in Italian). Dr. Colaneri is a Fellow of IFAC, the International Federation of Automatic Control and a Fellow of IEEE. He received the Certificate of Outstanding Service of IFAC. He served IFAC and IEEE Control Systems Society (CSS) in many capacities in journals and conferences. In particular, he was an Associate Editor of Automatica (certificate of outstanding service); a Senior Editor of IEEE Transactions on Automatic Control for eight years; a Senior Editor of the IFAC journal Nonlinear Analysis: Hybrid Systems. He has been the Chair of the Italian National Member Organization (CNR) for IFAC.
\end{IEEEbiography}
\end{document}